%% file: main.tex
\documentclass[preprint]{elsarticle}
\biboptions{sort&compress,merge}

\input{LZPaperSupport/packages.tex}
\input{LZPaperSupport/LZsym.tex}

\usepackage{comment}
\usepackage[english]{babel}
\usepackage{amsmath,,amssymb}
\usepackage[colorinlistoftodos]{todonotes}
\usepackage{multirow}
\usepackage{adjustbox}
\usepackage{comment}
\usepackage{float}
\usepackage{textcomp}
\usepackage[justification=justified]{subcaption}
\usepackage[justification=justified]{caption}
\usepackage[compact]{titlesec}
\usepackage{graphicx}
\usepackage{epstopdf}
\usepackage{color}
\usepackage{dcolumn}
\usepackage{hyperref}
\usepackage{xfrac}
\usepackage{threeparttable} 
\usepackage{enumitem}
\usepackage{placeins}
\begin{document}
\setlength{\tabcolsep}{3pt}


\title{A Liquid Scintillation Detector for Radioassay of Gadolinium-Loaded Liquid Scintillator for the LZ Outer Detector}

\author[ucsb]{S.J.~Haselschwardt\corref{cor1}}
\cortext[cor1]{Corresponding author}
\ead{sjh@physics.ucsb.edu}

\author[ucsb]{S.~Shaw}
\author[ucsb]{H.N.~Nelson}
\author[ucsb,lbnl]{M.S.~Witherell}
\author[bnl]{M.~Yeh}
\author[lbnl]{K.T.~Lesko}
\author[lbnl]{A.~Cole}
\author[ucsb]{S.~Kyre}
\author[ucsb]{D.T.~White}

\address[ucsb]{University of California (UC), Santa Barbara, Department of Physics, Broida Hall, Santa Barbara, CA 93106-9530, USA}

\address[lbnl]{Lawrence Berkeley National Laboratory (LBNL), 1 Cyclotron Road, Berkeley, CA 94720-8099, USA}

\address[bnl]{Brookhaven National Laboratory (BNL) P.O. Box 5000, Upton, NY 11973-5000, USA}

\begin{abstract}

We report on the design and performance of the LUX-ZEPLIN (LZ) ``Screener", a small liquid scintillator detector consisting of $\approx 23$ kg of LAB-based gadolinium-loaded liquid scintillator (GdLS) to be used in the LZ Outer Detector. The Outer Detector will be filled with 17.3 tonnes of GdLS and will surround the central liquid xenon time projection chamber of LZ. Its primary function will be to tag neutron events in the liquid xenon which could mimic a WIMP dark matter signal. To meet the deadtime requirements for the Outer Detector, the radioimpurity levels in the GdLS must be kept below $\lesssim0.07$ mBq/kg. This background level corresponds to a rate of $\approx50$ Hz above an energy threshold of 100 keV.

The Screener was operated in the ultra-low-background environment of the former LUX water shield in the Davis Laboratory at the Sanford Underground Research Facility for radioassay of the GdLS. Careful selection of detector materials and use of ultra-low-background PMTs allows the measurement of a variety of radioimpurities. The $^{14}\textrm{C}$/$^{12}\textrm{C}$ ratio in the scintillator is measured to be $(2.83\pm0.06\textrm{(stat.)}\pm0.01\textrm{(sys.)}) \times 10^{-17}$. Use of pulse shape discrimination allows the concentration of isotopes throughout the $^{238}\textrm{U}$, $^{235}\textrm{U}$, and $^{232}\textrm{Th}$ chains to be measured by fitting the collected spectra from $\alpha$ and $\beta$ events. We find that equilibrium is broken in the $^{238}\textrm{U}$ and $^{232}\textrm{Th}$ chains and that a significant portion of the contamination in the GdLS results from decays in the $^{227}\textrm{Ac}$ subchain of the $^{235}\textrm{U}$ series.

Predictions for the singles rate in the Outer Detector are presented. The rate from radioimpurities above 100 keV in the GdLS is estimated to be $97.9\pm6.4$ Hz, with $65.5\pm1.9$ Hz resulting from $\alpha$-decays.
\end{abstract}


\maketitle

\newpage

\section{\label{sec:Introduction}Introduction}

Searches for rare events caused by a variety of existing or conjectured particle and nuclear physics phenomena often feature auxiliary detectors that surround the principal detector in order to identify and permit the rejection of undesired events that can cause a background to the sought-after signal. Examples include Compton-suppression in high-sensitivity \Pgg-ray spectroscopy~\cite{Albert:1953,Landsberger:1996} and in searches for neutrinoless double-beta decay (\Dznbb)~\cite{Goulding:1984}; solar neutrino experiments~\cite{Hirata:1988ad,fukuda:2002uc}, and searches for dark matter~\cite{akimov:2010hr,ghag:2011jd}.

The LZ (LUX-ZEPLIN) experiment is a second generation direct dark matter detector~\cite{Mount:2017qzi,Akerib:2018lyp} under construction one mile underground (4300 meters water equivalent) in the Davis Laboratory of the Sanford Underground Research Facility (SURF) in Lead, South Dakota, USA~\cite{Heise:2015vza}. LZ uses a \SI{7}{\tonnel} central liquid xenon (LXe) target, arranged in a dual-phase time projection chamber (TPC), to seek evidence for nuclear recoils from a hypothesized galactic flux of Weakly Interacting Massive Particles (WIMPs). Two active detector elements surround the TPC: a layer of liquid xenon, the Xenon Skin (XS), optimized to detect \Pgg's, and the Outer Detector (OD), optimized to detect neutrons (\Pn's).

In this paper we describe studies undertaken at SURF with a small detector that we call the ``Screener'' to evaluate the liquid scintillator (LS) planned for use in the OD. A primary goal of these studies was to evaluate radioactive contamination levels of the LS, and determine whether it meets the requirements of the LZ experiment.

The LS used in large-mass experiments devoted to the detection of \Pgn's is one of the most radiopure materials known~\cite{Agostini:2017ixy,Gando:2014wjd} and would easily satisfy the radiopurity requirements for LZ. The purpose of the LS detector in LZ is, however, to detect \Pn's, so the requirements are similar to those for the detectors in \Pagne experiments, which are sensitive to the process $\Pagne+\Pp\to\Pep+\Pn$. The addition of gadolinium (Gd) to LS (GdLS) enhances the efficiency for \Pn detection~\cite{Piepke:1999db} and has been used successfully in several \Pagne experiments~\cite{Boehm:2001ik,An:2016ses,RENO:2015ksa,Abe:2014bwa}. However, since these experiments benefit from the high flux of \Pagne from nuclear reactors and have higher energy thresholds, the radiopurity of the GdLS was not a great concern. This study is designed to measure the radiopurity with the sensitivity needed for LZ. Other studies undertaken after this work was started~\cite{Fernandez:2016eux,Perez:2017} have found that isotopes from the \IUtTF chain can be present at a level far out of proportion to the usual relative abundance of \IUtTF to \IUtTe. That result had not been predicted, although in retrospect the enhanced level of \IPatTo and/or \IActtS in Gd is understandable ~\cite{hyde:1964v2}.

The GdLS mass of the detector described here is \SI{\approx23}{\kg}, and runs lasted roughly 10~days. Backgrounds were suppressed by the following techniques: 1) photomultiplier tubes (PMTs) with extremely low radioactivity (\SI{\approx1}{\mBq}/tube) designed for the LZ LXe TPC~\cite{akerib:2012da} were used to measure scintillation light; 2) the detector was operated inside the LUX shield of ultra-pure water~\cite{akerib:2012ys}; and 3) pulse-shape discrimination (PSD) was used to detect \Pga's with negligible background. These features enabled a single event sensitivity of \SI{\approx e-4}{\mBqkg} for \Pga's. Sensitivity for \Pgb's and \Pgg's is generally worse, since PSD is not useful in removing backgrounds. The typical sensitivity needed to assess whether a particular isotope might interfere with the LZ experiment's physics goals is \SI{\approx e-2}{\mBqkg}.

The power of the small detector described here enables it to perform a number of measurements relevant for low-background LS. The detector is easily sensitive to \ICof contamination at the \ICof/\ICot level of $10^{-19}$. Contamination with \ICof is a background to the detection of \Ipp neutrinos and other rare event physics~\cite{An:2015jdp,Agnes:2015qyz}. Additionally, the detector is capable of measuring deviations from secular equilibrium in the \IUtTe, \IUtTF, and \IThtTt chains.

\section{\label{sec:Requirements}Requirements of LZ}

\begin{figure}[h!]
\centering
\begin{subfigure}{0.52\textwidth}
\includegraphics[width=\textwidth]{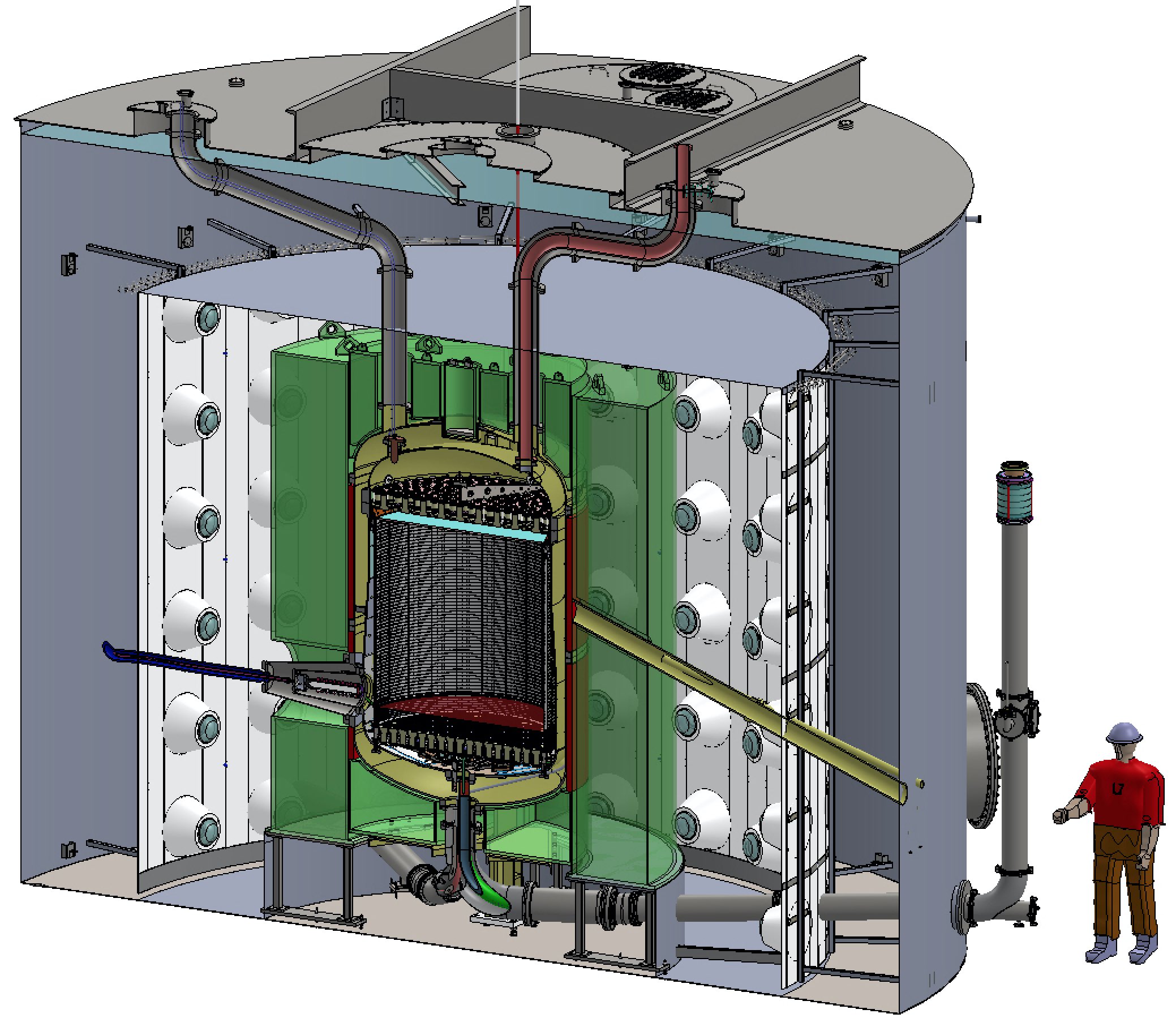}
\end{subfigure}
\begin{subfigure}{0.46\textwidth}
\includegraphics[width=\textwidth]{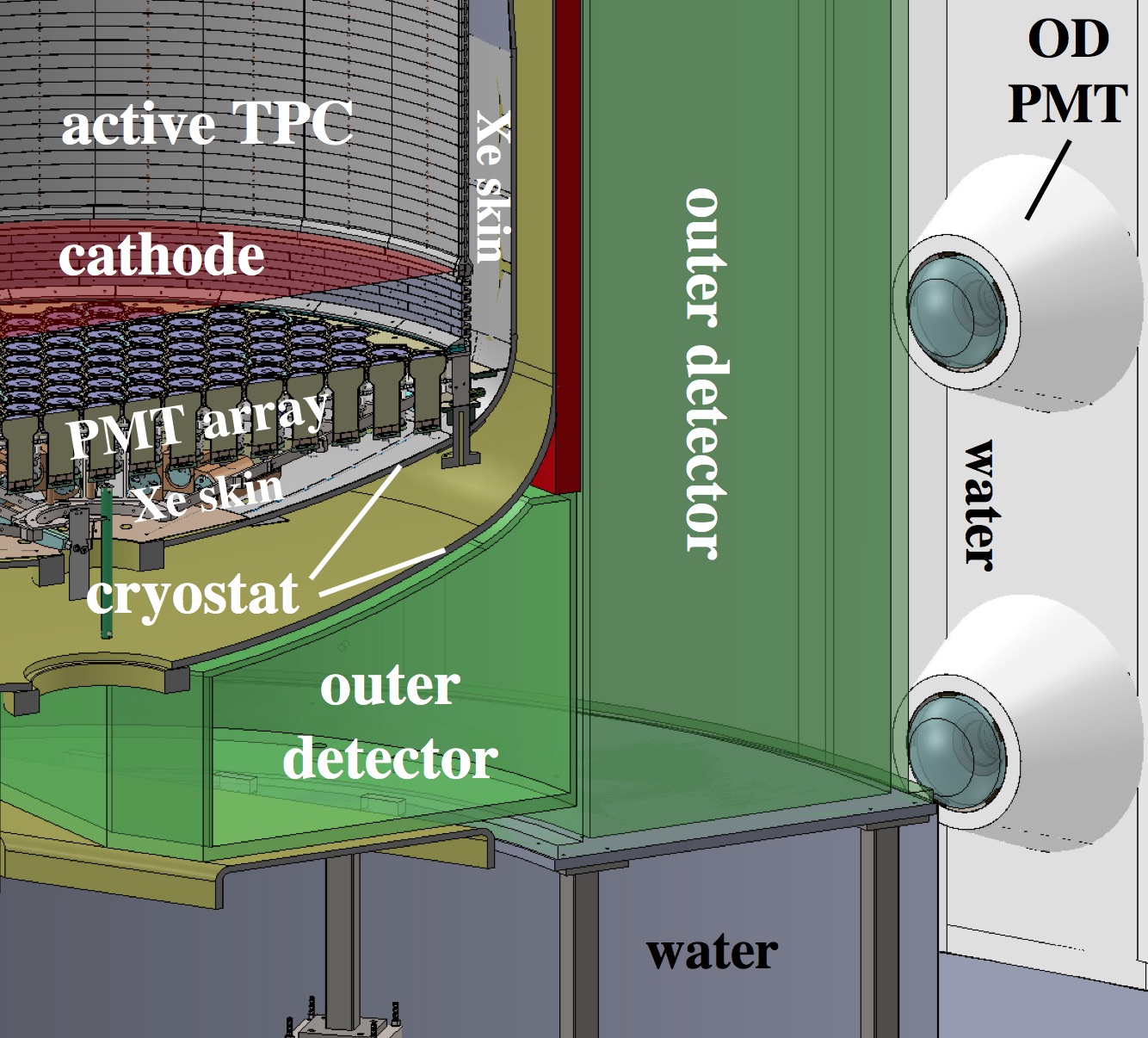}
\end{subfigure}
\caption{\small \textit{Left:} A cutaway drawing of the LZ detector in the underground water tank from Ref.~\cite{Akerib:2018lyp}. The OD acrylic tanks which hold the GdLS and surround the central TPC are shown in green. Light generated in the GdLS is collected in 120 PMTs which surround the tanks. A near-hermetic reflector system (white) aids in the OD light collection. \textit{Right:} A zoomed view of the detector also taken from Ref.~\cite{Akerib:2018lyp} showing details of the TPC internals, the XS region, and two OD PMTs. A full color version of this image is available online.\label{fig:LZ}}
\end{figure}

Two active detector elements surround the LZ TPC: the XS, a \SIrange{4}{8}{\cm} thick layer of liquid xenon, in the same cryostat as the TPC, and the OD, a \SI{\approx60}{\cm} layer of LS doped with \SI{0.1}{\percent} Gd by mass. A layer of ultra-pure water \SI{\approx200}{\cm} thick surrounds the assembly, contained in a stainless steel tank, to attenuate \Pn's and \Pgg's emitted from the Davis Laboratory walls. A drawing of the entire LZ detector system is shown in Fig.~\ref{fig:LZ}.

The signal sought in LZ is the recoil of a Xe nucleus in the TPC due to a collision with an impinging WIMP, accompanied by no energy depositions in the XS or the OD.

A variety of backgrounds to a WIMP signal arise from radioactive impurities in and near to the LXe TPC. An important class of backgrounds arise when a \Pgg or a \Pn is emitted due to a radioactive impurity in material on the perimeter of the LXe TPC, scatters once in the LXe TPC causing an energy deposit, and then exits. The XS has the principal power to veto \Pgg's, because of the small amount of material separating it from the LXe TPC, while the OD is designed to veto \Pn's. The LZ requirement is to achieve an efficiency for the detection of \Pn's which single-scatter in the LXe TPC of \SI{>95}{\percent}.

The main sources of \Pn's are the (\Pga,\Pn) process and fission, both caused by uranium and thorium impurities in material surrounding the TPC. Neutrons of kinetic energy \SI{\approx1}{\MeV} or higher can cause Xe recoils in the same energy window as those expected from a WIMP signal. The interaction length of \SI{1}{\MeV} \Pn's in LXe is about \SI{11}{\cm}, allowing a \Pn from the perimeter to enter the LXe TPC, scatter once, exit, and then penetrate to the OD. Collisions with hydrogen in the LS rapidly moderate the \Pn's kinetic energy to a near-thermal level of a fraction of an eV. The \Pn then diffuses until captured on a nucleus. In a predominantly hydrogenous material such as LS, the mean time to \Pn-capture on a proton is \SI{\approx220}{\us}, resulting in the emission of a \SI{2.2}{\MeV} \Pgg and a deuteron.

Loading of Gd in LS exploits two \Pn-capture properties of Gd. First, the thermal \Pn-capture cross section of natural Gd, \SI{4.3e4}{\barn}, is large, so addition of only \SI{0.1}{\percent} Gd by mass reduces the mean time to \Pn-capture to \SI{\approx30}{\mus}~\cite{An:2016ses}. Reduction of the time to \Pn-capture can reduce the rate of false vetoes caused by accidental activity. The LZ goal of a probability of false veto \SI{<5}{\percent} sets the maximum levels of radioactive impurities in and around the OD. Second, \Pn-capture on Gd is followed by a cascade of several \Pgg's which release a total energy of \SI{\approx8}{\MeV}. The probability of missing all of the several \Pgg's in this cascade is substantially lower than missing the single \SI{2.2}{\MeV} \Pgg from capture on hydrogen.

The acrylic vessels which hold the GdLS influence the OD performance. Engineering and logistical considerations dictate that the tanks be built in segments at the vendor on the surface and then transported down to the Davis Laboratory. Four main side vessels surround the LXe cryostat in quadrants that together form a cylinder \SI{375}{\cm} tall; six small tanks cover the top and bottom of the cryostat. The radial GdLS thickness of \SI{60}{\cm} cannot be made thinner due to manufacturing concerns; consideration of \Pn detection efficiency alone would permit thinner GdLS. The GdLS mass in the LZ OD is \SI{17.3}{\tonnesl}.

The \SI{2.5}{\cm} thickness of the inner side acrylic tank wall, facing the LXe cryostat, is required for structural integrity. Simulations indicate that \SIrange{5}{10}{\percent} of \Pn's emitted from the LXe TPC capture in the acrylic inner walls of the GdLS tanks, with a mean capture time characteristic of capture on protons: \SI{\approx220}{\mus}. To allow detection of these \Pn's, a time window for \Pn detection of \SI{500}{\mus} is set; the start of the time window will be a signal from the LXe TPC.

To achieve a probability of a false veto \SI{<5}{\percent}, the singles rate in the OD system must be lower than \SI{100}{\Hz} above an energy threshold of \SI{100}{\keV}. Measurements and simulations predict a rate from \Pgg's emitted from the Davis Laboratory walls which penetrate the water shield, as well as from nearby LZ materials, of \SI{50}{\Hz}, allowing at most \SI{50}{\Hz} to arise from radioactive impurities in the GdLS.

Radioactive impurities in the GdLS consist of generally common isotopes and more unusual isotopes which enter with gadolinium. Common isotopes include \ICof, \IKreF, \IKfz, and the isotopes in the \IUtTe and \IThtTt chains. Unusual isotopes that enter with \IGd include \ILaoTe, \ISmofS, \IGdoFt, \ILuoSs, and isotopes in the \IUtTF chain. Approximately 40 isotopes produce energy deposits above a threshold of \SI{100}{\keV}, resulting in a specific activity for an average isotope that should not exceed \SI{0.07}{\mBqkg}. Natural gadolinium contains \SI{0.2}{\percent} \IGdoFt by weight, which contributes \SI{1.6}{\mBqkg} in GdLS with \SI{0.1}{\percent} gadolinium by mass. The signal from the \SI{2.2}{\MeV} \Pga from \IGdoFt is heavily quenched and emits light similar to that of a \SI{100}{\keV} \Pgb signal.

The abundance of other radioactive impurities can be influenced by chemical processing and purification of the components of GdLS. The specific activity levels reported by a variety of experiments which have used LS, as well as by experiments that use gadolinium, converted to the expectation in the GdLS described in this publication, are given in Tab.~\ref{tab:worldLS}. Many of the specific activities exceed \SI{0.07}{\mBqkg}.

\begin{table}[h]
\centering
\footnotesize
\caption{\small Summary of specific activities in units of mBq/kg, converted to a kilogram of LZ GdLS. To meet the LZ OD specifications, an average specific activity \SI{\leq0.07}{\mBqkg} is needed. \label{tab:worldLS}}
\begin{tabular}{l c c c c c c} \hline
\multirow{2}{*}{\textbf{Isotope}} & \textbf{Large Detectors} & \multicolumn{2}{c}{\textbf{GADZOOKS! Gd~\cite{Perez:2017}}} & \multicolumn{2}{c}{\textbf{Boiko Gd~\cite{Boiko:2017mmh}}} \\
 & \textbf{Highest} & \textbf{Lowest} & \textbf{Highest} & \textbf{Purified} & \textbf{Raw} \\ \hline
\vphantom{\large\ICof}\ICof & 5000~\cite{Agnes:2015qyz} &  &  &  & \\
\IKfz & 6.3~\cite{Monzani:2005hxa} & \num{<1.9e-3} & \num{0.25+-0.01} & $<0.04$ & $<0.09$ \\
\IKreF & \num{0.88+-0.02}~\cite{Keefer:2009} &  & &  &  \\
\IUtTeL/\IPotoz & \num{0.20+-0.05}~\cite{An:2016ses} &  & &  &  \\
\IUtTeL/\IBitoz & \num{0.07+-0.01}~\cite{Keefer:2009} &  & &  &  \\
\IUtTeL/\IPbtoz & \num{0.06+-0.01}~\cite{Keefer:2009} &  & &  &  \\
\IThtTtE &  & \num{<3.9e-4} & \num{2.12+-0.02} & $<0.005$ & \num{0.09+-0.01} \\
\IThtTtL & \num{\approx8e-3}~\cite{An:2016ses} & \num{5.8+-4e-4} & \num{0.97+-0.01} &  &  \\
\IUtTeE  &  & $<0.013$ & \num{1.9+-0.1} & $<0.8$ & $<1.3$ \\
\IUtTeM  & \num{\approx4.5e-4}~\cite{An:2016ses} & \num{<3.8e-4} & \num{0.143+-0.004} & $<0.01$ & $<0.009$ \\
\IUtTFE  & & \num{<7.7e-4} & $<0.22$ & $<0.01$ & \num{0.11+-0.01} \\
\IUtTFL  &  \num{\approx0.01}~\cite{An:2016ses} & \num{<1.0e-3} & \num{4.7+-0.1} & \num{2.2+-0.1} & \num{1.6+-0.1} \\
\ILaoTe  &  & \num{<2.9e-4} & \num{1.32+-0.03} & \num{0.030+-0.003} & \num{0.014+-0.002} \\
\ILuoSs  &  & \num{4.4+-1.7e-4} & \num{1.09+-0.01} & \num{0.035+-0.003} & \num{0.038+-0.004} \\
\hline
\end{tabular}
\end{table}

The two uranium decay chains and the thorium decay chain each contribute $\approx10$ unstable isotopes. A useful starting point is to assume that the isotopes in these chains are in secular equilibrium, with all isotopes within one chain, from the head  (\IUtTe, \IUtTF, and \IThtTt, respectively) to a step prior to the stable final isotope (\IPbtzs, \IPbtzS, and \IPbtze, respectively) contributing a rate equal to all other isotopes within the respective chain. Violations of secular equilibrium make the definition of subchains useful~\cite{ malling:2013jya}.

\begin{figure}[t]
\centering
\includegraphics[width=0.98\textwidth]{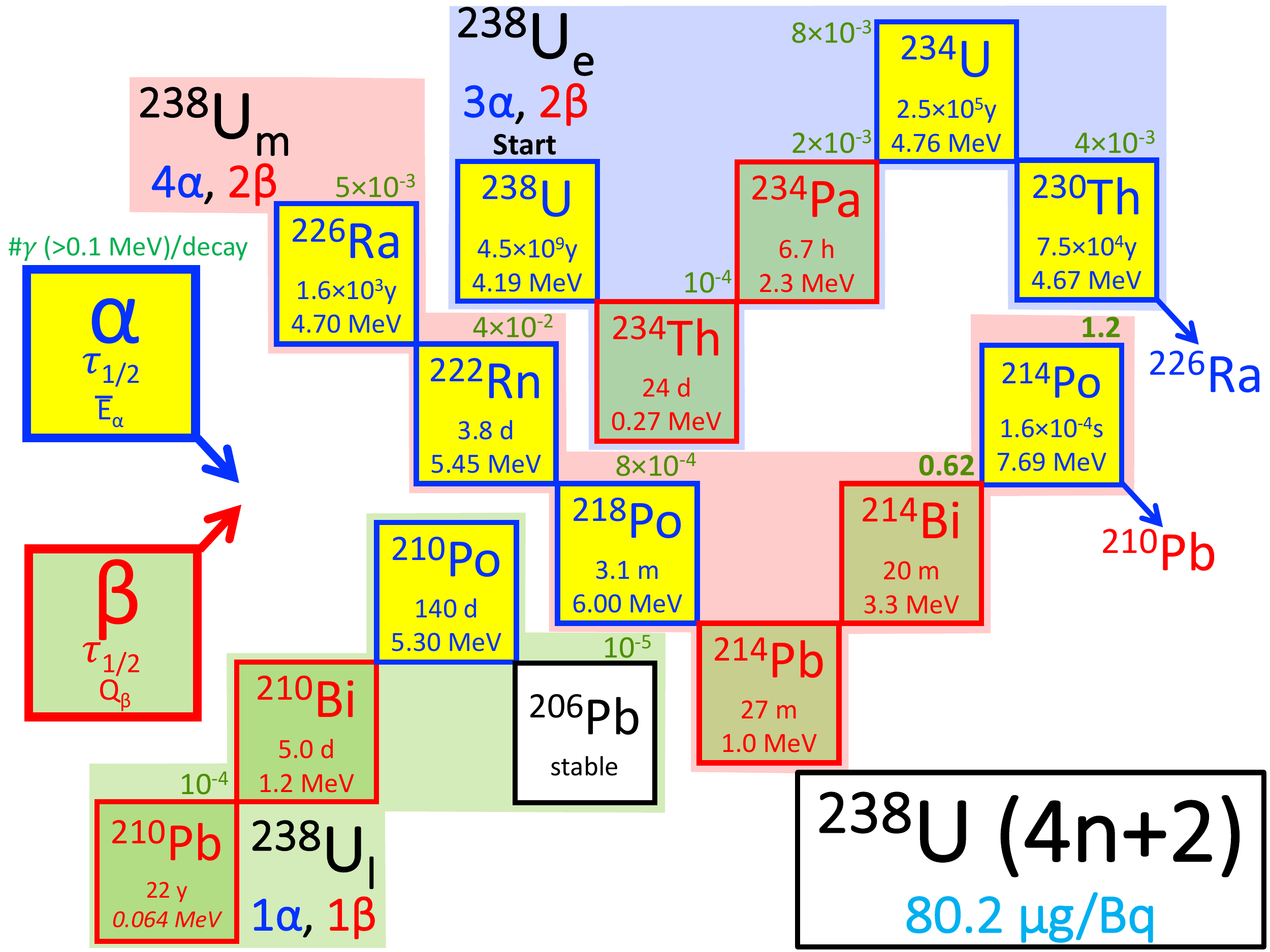}
\caption{\small Definition of subchains for \IUtTe. The 14 principal radioactive isotopes in the \IUtTe decay chain are each shown in a box, with the half-life and energy released into \Pga's and into \Pgb's. The branching ratio into clear (\SI{>100}{\keV}) \Pgg's is shown in bold above the daughter, while for weak \Pgg emitters, the largest branching ratio is shown. The one isotope with an energy release below the OD threshold, \IPbtoz, is not included in OD rate analyses or in the count of \Pgb decays shown. A full color version of this image is available online.\label{fig:UtTesub}}
\end{figure}

The subchains defined for \IUtTe are shown in Fig.~\ref{fig:UtTesub}. Of the five isotopes in the early \IUtTeE subchain, three are long-lived, and are chemically similar, while the other two re-equilibrate rapidly after chemical processing.  The six isotopes in the middle \IUtTeM subchain equilibrate rapidly with the chemically distinct \IRatts after processing. Emanation of \IRnttt from surfaces contributes to \IUtTeM.  The level of \IUtTeM is often measured in LS by the distinct \Pgb-\Pga pulse pair from \IBitof and \IPotof. Of the three radioactive isotopes in the late \IUtTeL subchain, the \Pgb from \IPbtoz has an energy below the OD threshold of \SI{100}{\keV}.  Surfaces contain \IPbtoz deposited by environmental \IRnttt and frequently contribute \IBitoz and \IPotoz at rates higher than the other \IUtTe subchains.

\begin{figure}[h]
\centering
\includegraphics[width=0.98\textwidth]{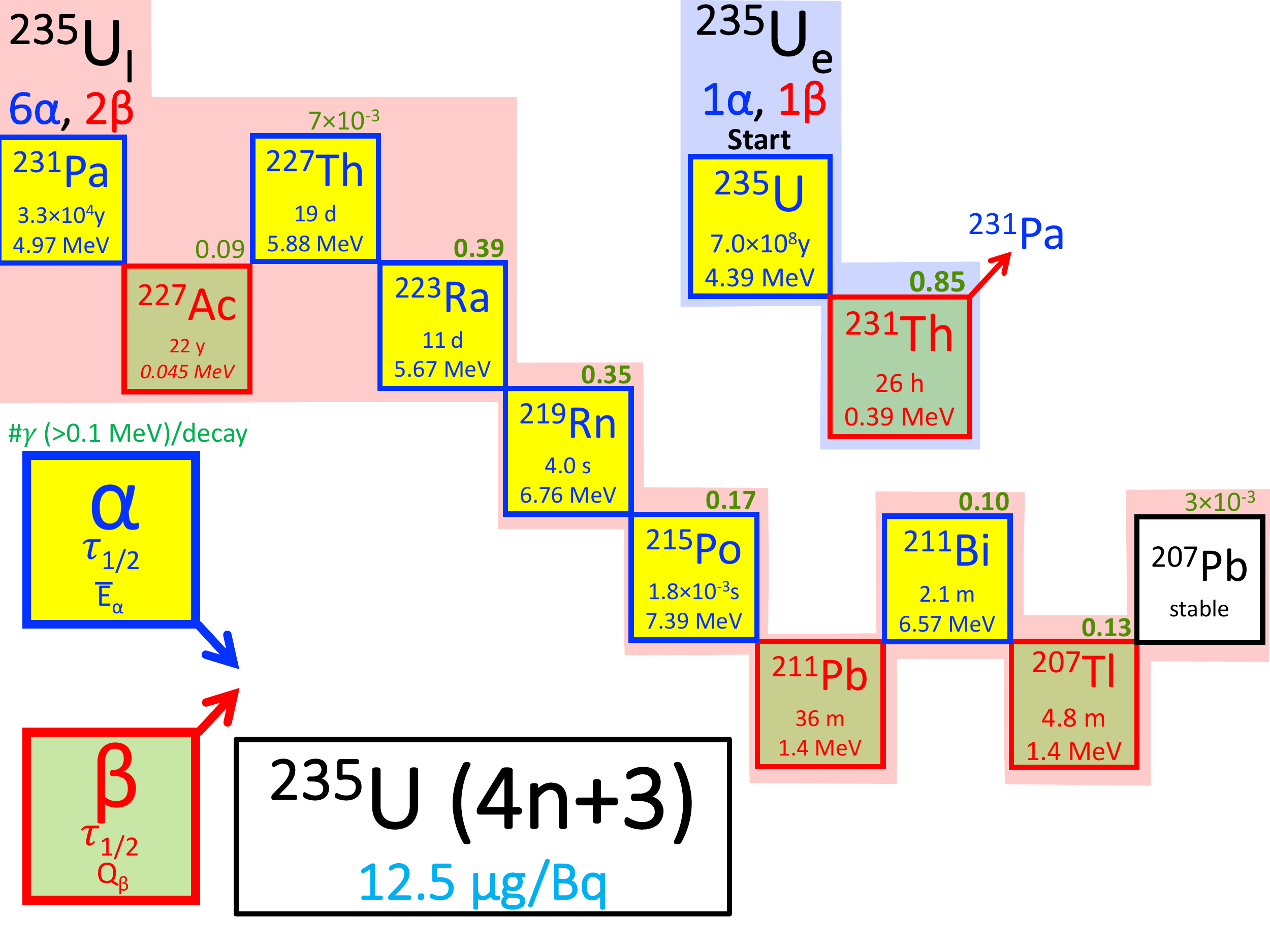}
\caption{\small  Definition of subchains for \IUtTF. The 11 principal radioactive isotopes in the \IUtTF decay chain are each shown in a box, with the half-life and energy released into \Pga's and into \Pgb's.  The branching ratio into clear (\SI{>100}{\keV}) \Pgg's is shown in bold above the daughter, while for weak \Pgg emitters, the largest branching ratio is shown. The one isotope with an energy release below the OD threshold, \IActtS, is not included in OD rate analyses or in the count of \Pgb decays shown. A full color version of this image is available online.\label{fig:UtTFsub}}
\end{figure}

The subchains defined for \IUtTF are shown in Fig.~\ref{fig:UtTFsub}. The early \IUtTFE subchain contains two isotopes: the long-lived head of the chain, \IUtTF, and the short-lived \IThtTo, which rapidly comes to equilibrium with \IUtTF after chemical processing.  The nine isotopes in the late \IUtTFL subchain start with the long-lived \IPatTo, which is chemically distinct from uranium and not necessarily suppressed by chemical processes that suppress uranium.  While \IActtS is long-lived enough to re-establish equilibrium quickly after chemical processing, its chemistry may be similar enough to that of protactinium to stay in equilibrium following chemical processing.  Both protactinium and actinium are chemically similar to gadolinium. The \Pgb from \IActtS has an energy below the OD threshold of \SI{100}{\keV}. The other seven isotopes in \IUtTFL rapidly re-equilibrate after chemical processing.  The level of \IUtTFL can be measured in LS by the distinct \Pga-\Pga pulse pair from \IRnton and \IPotoF. 

\begin{figure}[h]
\centering
\includegraphics[width=0.98\textwidth]{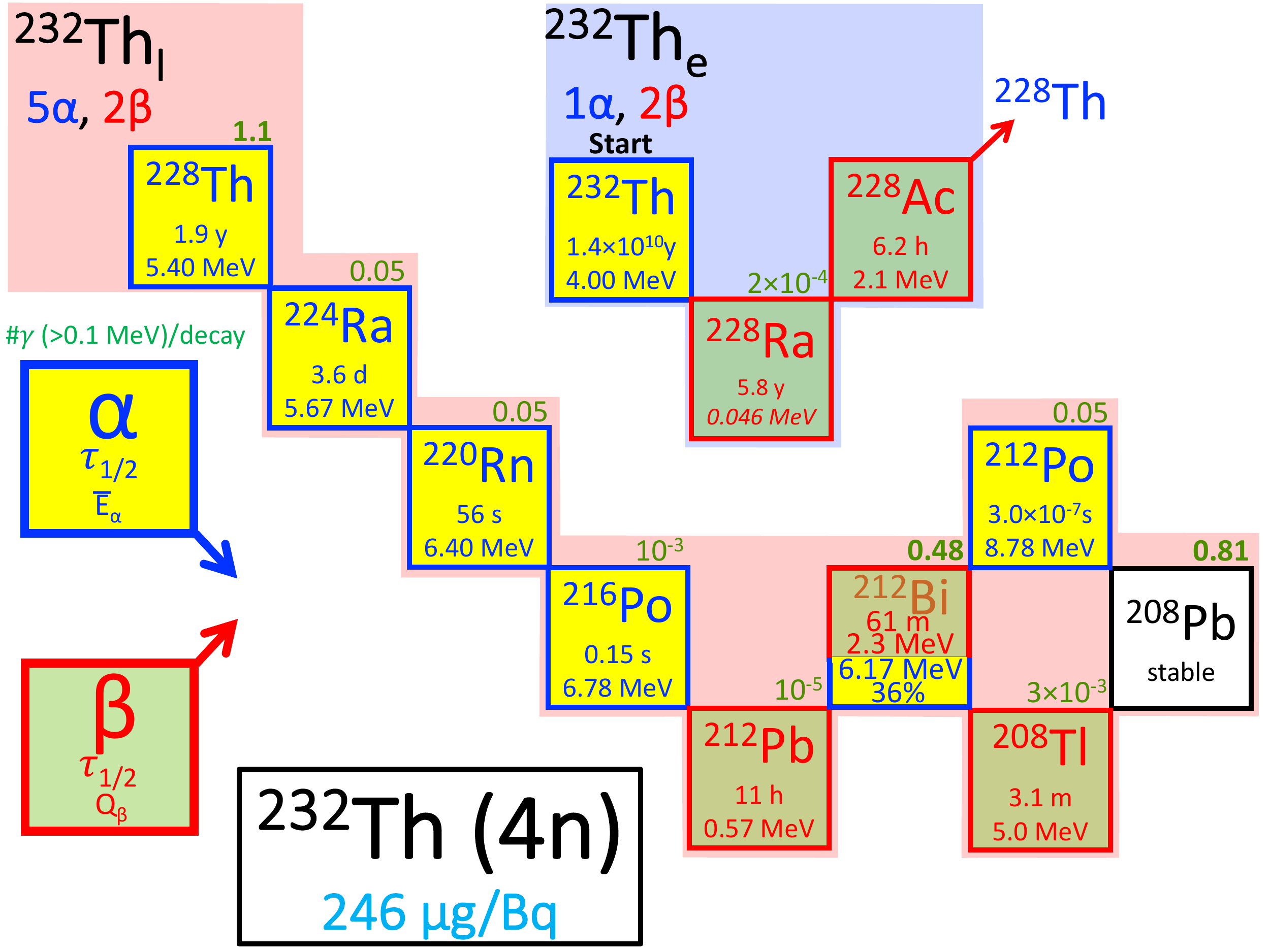}
\caption{\small Definition of subchains for \IThtTt. The 11 principal radioactive isotopes in the \IThtTt decay chain are each shown in a box, with the half-life and energy released into \Pga's and into \Pgb's. The branching ratio into clear (\SI{>100}{\keV}) \Pgg's is shown in bold above the daughter, while for weak \Pgg emitters, the largest branching ratio is shown. The one isotope with an energy release below the OD threshold, \IRatte, is not included in OD rate analyses or in the count of \Pgb decays shown. A full color version of this image is available online.\label{fig:ThtTtsub}}
\end{figure}

The subchains defined for \IThtTt are shown in Fig.~\ref{fig:ThtTtsub}. Of the three isotopes in the early \IThtTtE subchain, only the head isotope is long-lived, while the other two re-equilibrate rapidly after chemical processing. The \Pgb from \IRatte has an energy below the OD threshold of \SI{100}{\keV}. There are eight distinct radioactive isotopes in the late \IThtTtL subchain, but weighting by the branching ratios of \IBitot reduces the count to seven. The isotopes in \IThtTtL may rapidly re-equilibrate with \IThtte after chemical processing, and are influenced by emanation of \IRnttz from surfaces. The level of \IThtTtL is often measured in LS by the distinct \Pgb-\Pga pulse pair from \IBitot and \IPotot.

The isotope with the highest specific activity in Tab.~\ref{tab:worldLS} is \ICof. Use of underground hydrocarbon sources for the LS chemicals results in \ICof specific activities near \SI{1}{\mBqkg}, while use of biogenic hydrocarbon sources caused the higher value in the table. The \ICof \Pgb-decay spectrum endpoint is \SI{156.5}{\keV}, with \SI{8}{\percent} of single pulses exceeding \SI{100}{\keV}. An increase of energy threshold can suppress single pulses from \ICof, but multiple pulses from pileup of \ICof decays during the planned LZ time window of \SI{500}{\mus} can still be troublesome. The surveying technique of accelerator mass spectrometry has a sensitivity to \ICof/\ICot of \num{e-15}, corresponding to a specific activity of \SI{100}{\mBqkg}, which is not sufficient for the LZ OD. The desire to measure \ICof/\ICot with the required sensitivity is one of the principal motivations for building the Screener.

Isotopes in the \IUtTF chain have been detected in gadolinium for GADZOOKS!~\cite{Fernandez:2016eux} and also in GdLS at the level of \SI{0.01}{mBq/kg}~\cite{An:2016ses}. In Tab.~\ref{tab:worldLS}, the specific activity per kg LZ GdLS is extrapolated from various samples studied in Ref.~\cite{Perez:2017}. The lowest and highest values in Ref.~\cite{Perez:2017} arise from variation among the samples obtained from different suppliers. Tab.~\ref{tab:worldLS} also shows the specific activities in LZ GdLS from the gadolinium evaluated in Ref.~\cite{Boiko:2017mmh}, both before and after purification. The highest \IUtTF levels from both \IGd studies exceed the specification for LZ GdLS.

Two radioactive isotopes are not included in Tab.~\ref{tab:worldLS}. The first is \IGdoFt, which is unavoidable when using natural gadolinium and has \SI{\approx 60}{\percent} probability of exceeding a light output equivalent to a \SI{100}{\keV} \Pgb pulse. A rate in the OD of \SI{\approx17}{\Hz} is expected from \IGdoFt. \ISmofS emits an \Pga of energy \SI{2.3}{\MeV}, which will have a \SI{\approx 70}{\percent} chance of exceeding the light output of a \SI{100}{\keV} \Pgb pulse. The Screener is able to detect both of these \Pga emitters.

\section{\label{sec:DetectorDescription}The Screener Detector System}

\subsection{Physical Detector}
Detector materials were chosen to minimize radioactivity near the LS volume, and all materials were screened by high-purity germanium (HPGe) \Pgg-counting~\cite{Mount:2017iam}. Plastic rather than metal components were favored to suppress radioimpurities.

\begin{figure}[t]
\centering
\begin{subfigure}{0.4\textwidth}
\includegraphics[height=165pt,keepaspectratio]{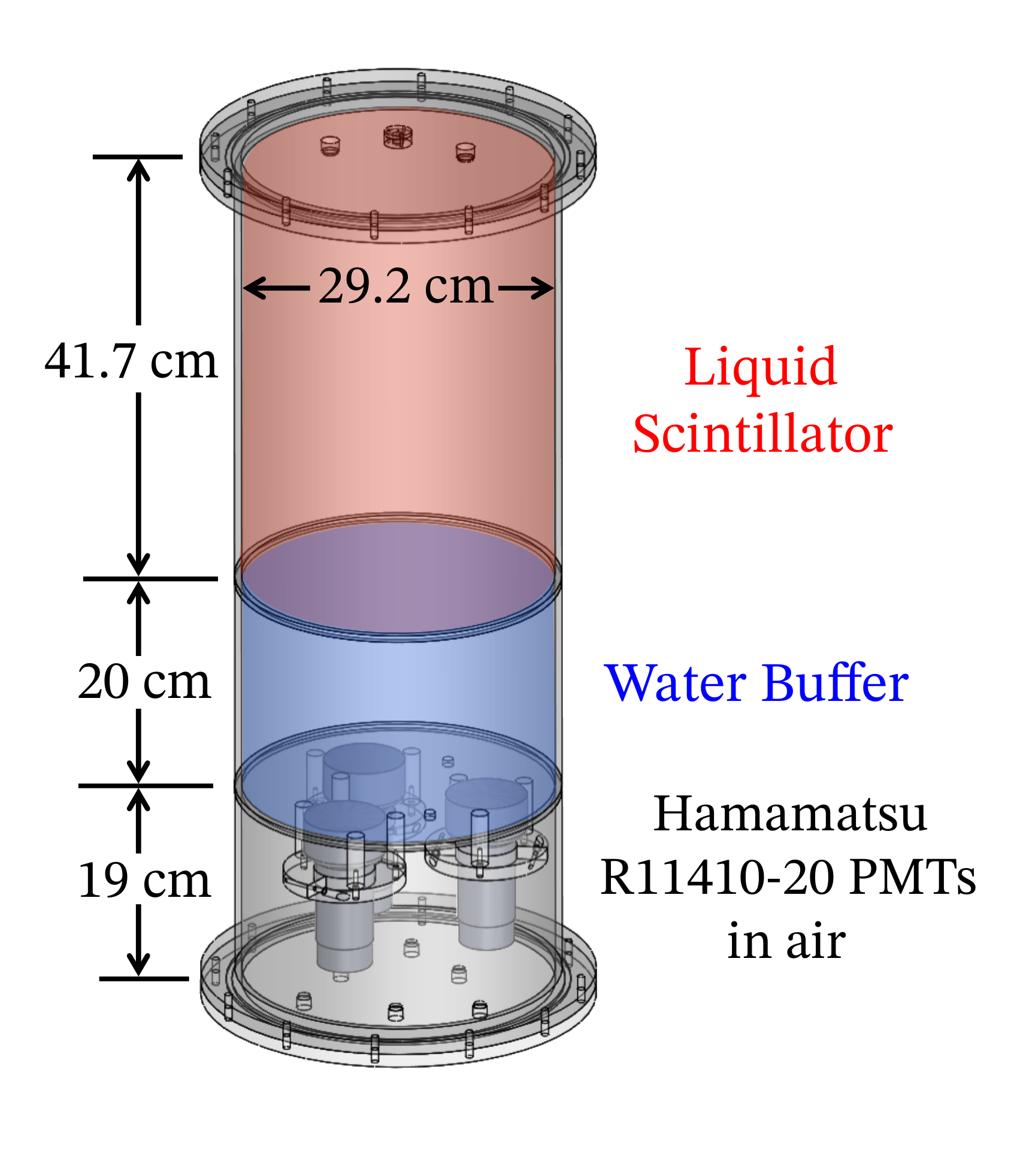}
\end{subfigure}
\begin{subfigure}{0.28\textwidth}
\includegraphics[height=165pt,keepaspectratio]{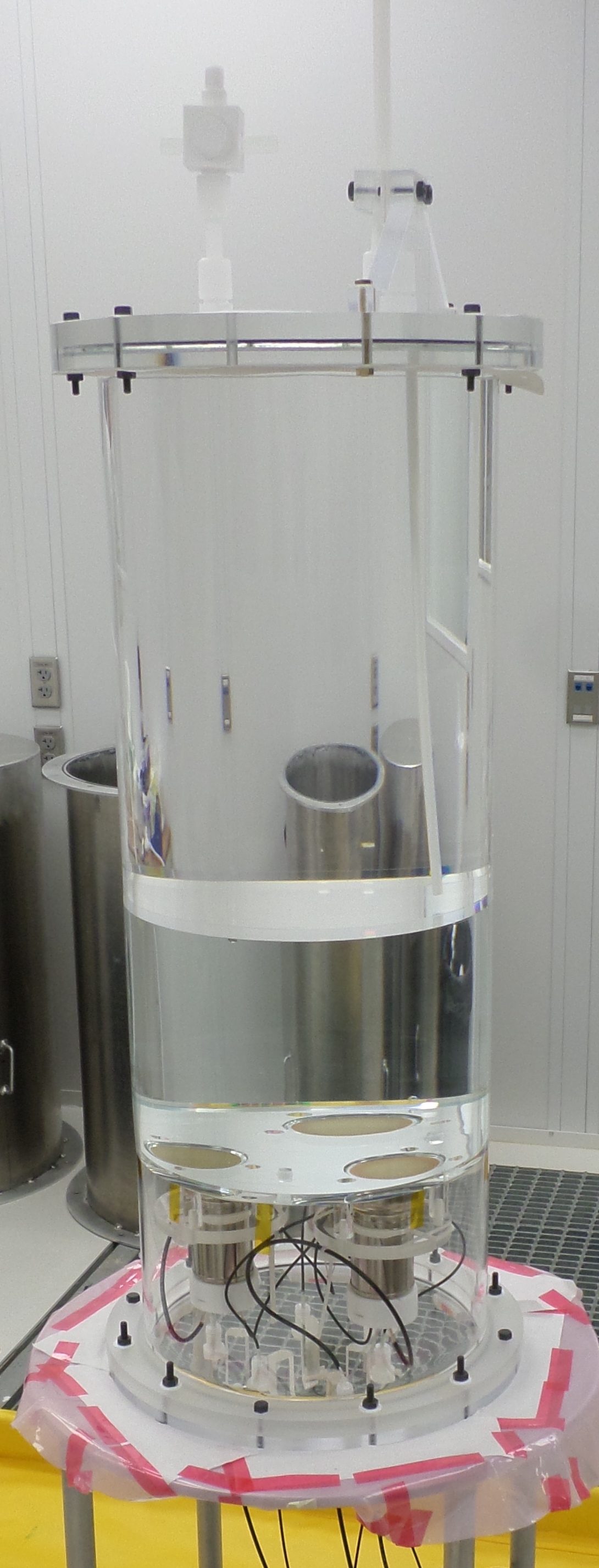}
\end{subfigure}
\begin{subfigure}{0.28\textwidth}
\includegraphics[height=165pt,keepaspectratio]{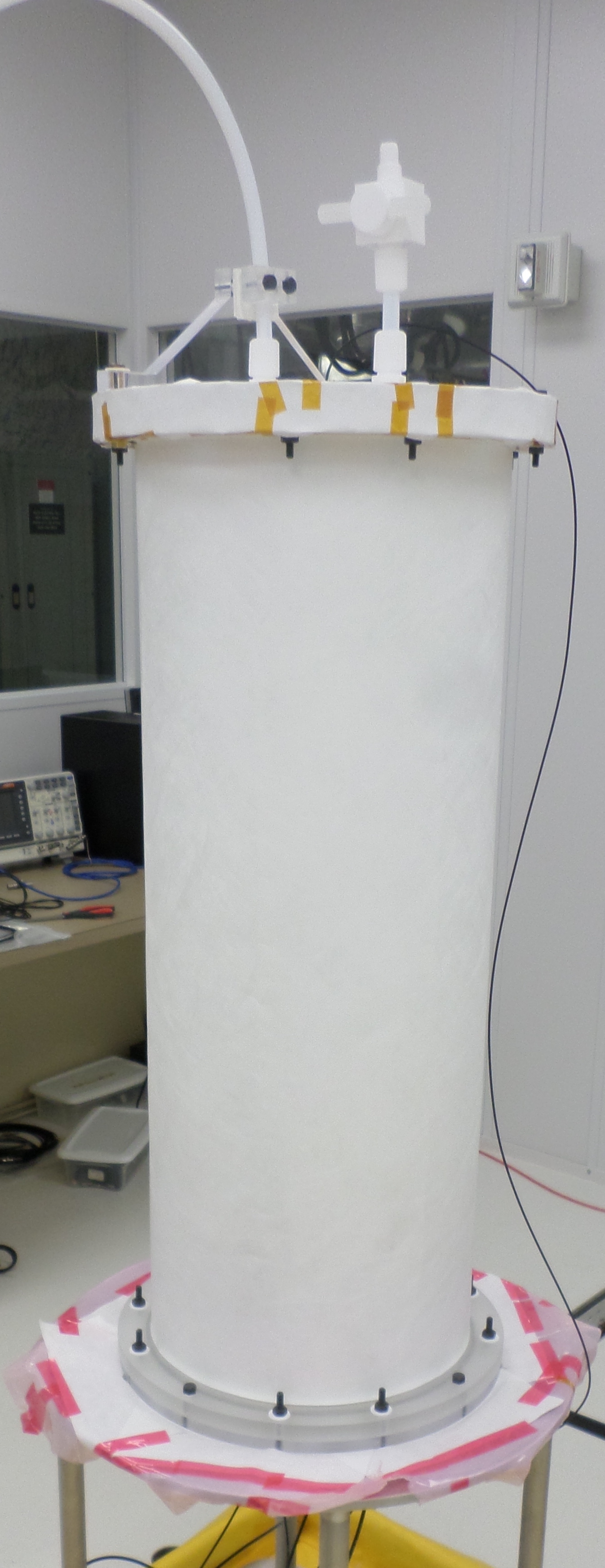}
\end{subfigure}
\caption{\small \textit{Left:} Diagram of the detector vessel. \textit{Middle:} The detector in the Davis Laboratory cleanroom with PMTs mounted, water and LS chambers filled. \textit{Right:} The detector wrapped with Tyvek reflector before deployment. A full color version of this image is available online.\label{fig:screener}}
\end{figure}

The detector is comprised of a clear UV-transparent acrylic tube of inner diameter \SI{29.2}{\cm} and wall thickness \SI{0.64}{\cm} segmented along its length into three chambers, fabricated by Reynolds Polymer Technology, Inc.~\cite{Reynolds:2018}. The partitions between chambers are \SI{0.64}{\cm} thick, and the chamber lids are \SI{1.3}{\cm} thick. The top chamber is \SI{41.7}{\cm} long and has a capacity of \SI{28}{\liter}, so it can hold up to \SI{24}{\kg} of LS. The middle chamber is \SI{20}{\cm} long and is filled with \SI{14.3}{\kg} of distilled, deionized water that shields the LS from \Pgg-rays generated in the PMTs. The bottom chamber is \SI{19}{\cm} long, filled with air, and holds the three low-background PMTs. The PMTs face the LS and are coupled to the acrylic wall between their chamber and the water volume using optical grease. The chambers on each end of the vessel are sealed by tightening 12 brass bolts and compressing an O-ring between the vessel flange and the chamber lid. The assembly is wrapped in \SI{262}{\micro\meter}-thick 1085D Tyvek~\cite{Dupont:2018,MaterialConcepts:2018}, which has a reflectivity \SI{\gtrsim95}{\percent} over the range of GdLS light emission of \SIrange{350}{550}{\nm}~\cite{Janecek:2008,Janecek:2012,Yu:2012zzc}. A diagram and two photographs of the detector are shown in Fig.~\ref{fig:screener}.

The three PMTs are \SI{7.6}{cm} Hamamatsu R11410-20, which have radioimpurity levels \SI{\approx1}{\mBq\per PMT} for the U/Th chain and \SI{12+-2}{\mBq\per PMT} for \IKfz\cite{Aprile:2015lha}. The R11410-20 has a quantum efficiency of \SI{\approx35}{\percent} at a wavelength of \SI{420}{\nm}, the peak of GdLS emission, and resolves single photoelectrons (phe)~\cite{akerib:2012da}.

Two fill ports on the LS chamber lid allow for filling and venting of the chamber. A flexible line of concentric PTFE tubing is left on one of the ports during deployment. The inner tubing is converted to a rigid PTFE tube that extends into the bulk of the LS volume allowing nitrogen gas carrying \IRnttz to be bubbled through the scintillator for calibration.

The detector was deployed in the former LUX water tank~\cite{akerib:2012ys} just after the LUX detector was removed. The water shields from \Pgg's emitted from uranium, thorium, and potassium in the Davis Laboratory walls. The water tank height is \SI{591}{cm}, and its radius is \SI{381}{\cm}. An inverted steel pyramid under the water tank of maximum thickness \SI{30.5}{\cm} and diameter \SI{500}{\cm} provides additional shielding from the rock in the floor. The stainless steel LUX detector stand was present in the water tank during the Screener deployment.

\begin{figure*}[t]
\centering
\includegraphics[width=0.98\textwidth]{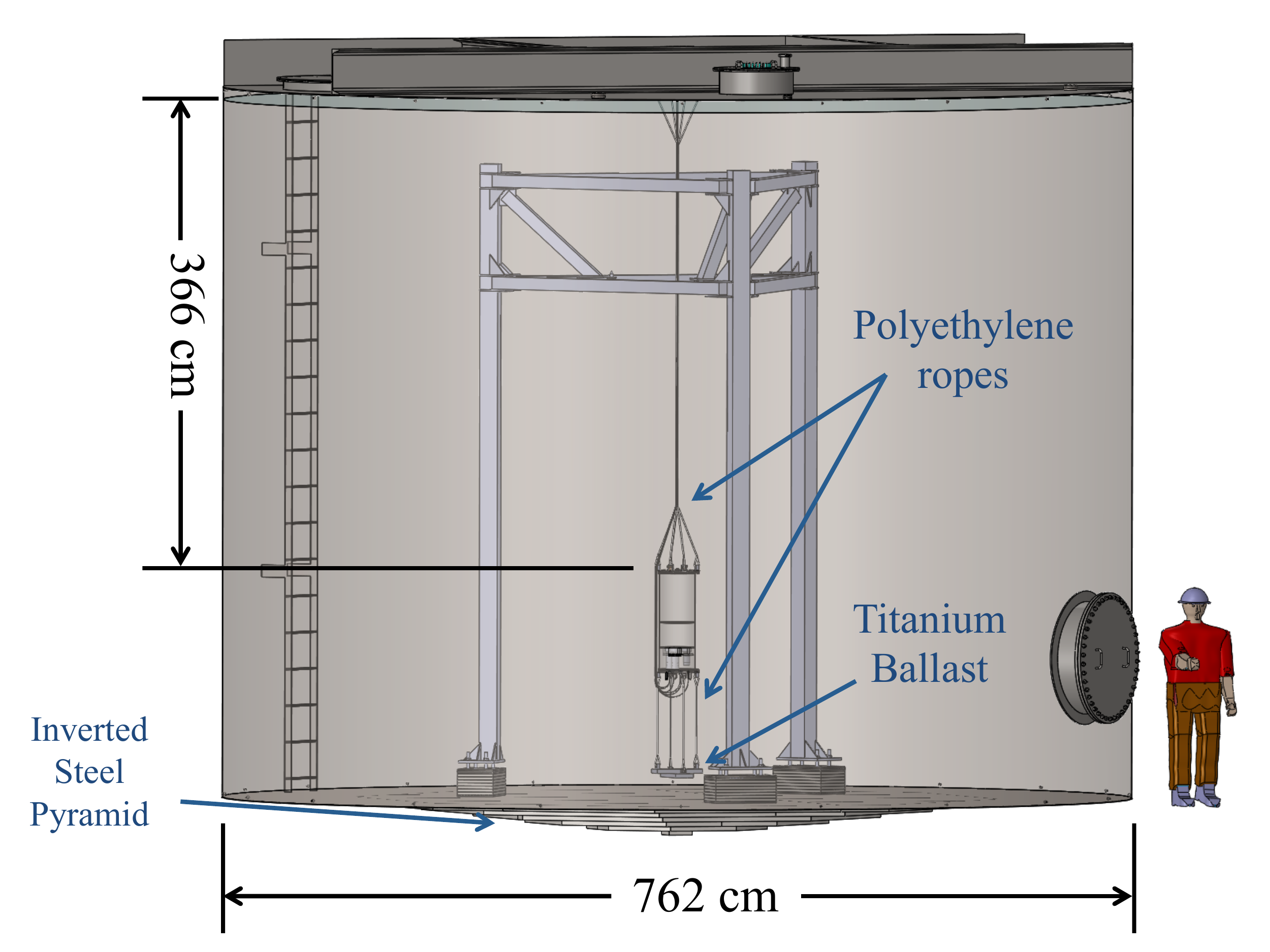}
\caption{\small Deployment of the Screener in the water tank.  The stand for the LUX detector is visible along with the inverted steel pyramid underneath the water tank floor. A full color version of this image is available online.\label{fig:screener_in_tank}}
\end{figure*}

Polyethylene ropes suspended the detector assembly inside the water tank, as shown in Fig.~\ref{fig:screener_in_tank}. A \SI{20}{\kg} ballast of ultra-pure titanium~\cite{Akerib:2017iwt} was suspended \SI{\approx80}{\cm} from the bottom of the vessel to counteract the vessel buoyant force and to provide stability.

The vertical position in the water tank was chosen to minimize the rate from external \Pgg's originating from the LUX detector stand and from the Davis Laboratory walls. The \Pgg's from \IUtTe, \IThtTt, \IKfz, and \ICosz decays were described with a \textsc{Geant4}-based~\cite{agostinelli:2002hh} simulation package. At this location, there was \SI{160}{\cm} (\SI{366}{\cm}) between the bottom (top) of the detector scintillator volume and the bottom (top) of the water volume. The estimated stand and wall contribution to the Screener rate is \SI{0.57+-0.03}{\mHz} above an energy deposit threshold of \SI{100}{\keV}~\cite{davisGammasPaper:2018}.

The resistivity and dissolved oxygen in the water was monitored during runs. An earlier measurement of radon in the nitrogen used to purge the water tank found a concentration of \SI{0.054+-0.026}{\Bq/\meter^3} while the dissolved oxygen was below the \SI{1}{\ppb} sensitivity of the oxygen monitor. Simulations predict a rate of \SI{3.6+-1.7}{\mHz} above \SI{100}{\keV} from \IRnttt decays in the water at this concentration.

The PMT signal and high voltage (HV) were provided via \SI{13.7}{\meter} long RG174 and RG58 coaxial cables, respectively. Each cable was potted with epoxy into a water-tight feedthrough in the PMT chamber lid, secured to the Screener vessel's top and bottom flanges and routed out of the water tank through a light-tight port. The cables were then connected to the HV power supply and data acquisition system (DAQ) in a nearby electronics rack. Simulations indicate that the cables are the single largest source of background from construction materials, their contribution being \SI{54}{\percent} of the total background rate. Screening of the cables had not been completed before deployment of the detector.

\subsection{Electronics and Data Acquisition}

The PMT signals went into charge sensitive amplifiers with high (40$\times$) and low (4$\times$) gain outputs for each input channel. Shaped and amplified waveforms were then digitized at \SI{100}{\MHz} by a SkuTek DDC-10~\cite{Skutek:2018} digitizer and stored for off-line processing. Waveform digitization was triggered on the condition of coincident signals of greater than \SI{\approx1.5}{\phe} in all three PMTs within a \SI{50}{\ns} time window. In this way, each PMT's $\mathcal{O}(\textrm{kHz})$ dark rate does not cause a significant number of triggers ($<1$~event in a 10~day run).

The digitized window for each trigger started \SI{2}{\micro\second} \emph{before} the trigger and extended after the trigger by \SI{79.9}{\micro\second}. A subsequent trigger was re-enabled after writing the event
to disk, which took on average \SI{30}{\ms}. 
The reconstruction of pulse pair rates
when the second decaying nucleus has
lifetime comparable to the digitization
window or the re-enable time interval
must account for deadtime.

The data were processed off-line by a pulse finding algorithm. The recorded waveforms were scanned for pulses and basic quantities such as the start and end times, area, and height were computed for each pulse found.

\section{\label{sec:LSprops}Liquid Scintillator}

The LS solvent is linear alkylbenzene (LAB), chosen largely for its high flash point of \SI{130}{\celsius}, making LAB combustible, not flammable~\cite{national2018nfpa}. This simplifies safety mitigations required in underground laboratories. The other components of GdLS~\cite{Ding:2008zzb} are given in Tab.~\ref{tab:GdLS}. The LAB is purchased from CEPSA~\cite{CEPSA:2018}, and is derived from underground sources, to minimize \ICof content~\cite{Alimonti:1998rc}.

All components of the GdLS are purified~\cite{Ding:2008zzb,Yeh:2010zz} with the exception of the bis-MSB (1,4-bis( 2-methylstyryl )benzene) because of the extremely small quantity used in GdLS. The LAB solvent is purified by distillation, while the PPO (2,5-diphenyloxazole) is cleaned using recrystallization and water extraction. The \IGd compound is purified by pH-controlled partial hydrolysis, during which \IGd remains in solution while certain radioactive actinides such as uranium and thorium are precipitated out. Actinium itself, however, tends to remain in solution with \IGd~\cite{Yeh:2010zz}, resulting in an out-of-equilibrium level of \IActtS.

\begin{table} [ht]
\caption{Chemical components in \SI{1}{\liter} of GdLS.}
\centering
\begin{tabular} 
{c c >{\centering}m{5em} c} 
\hline
    {\bfseries Acronym} &
    {\bfseries Molecular Formula} &
    {\bfseries Molecular Weight (g/mol)} & 
    {\bfseries Mass (g)} \\
\hline
\vphantom{\Large L}LAB & \CCHLAB & 234.4 & 853.55 \\
PPO & \CPPO & 221.3 & 3.00 \\
bis-MSB & \CbisMSB & 310.4 & 0.015 \\
TMHA & \CTMHAM & 157.2 & 2.58 \\ 
Gd & \IGd & 157.3 & 0.86 \\ \hline
\vphantom{\Large G}
GdLS & \CGdLS & 233.9 & 860.0 \\ \hline
\end{tabular}
\label{tab:GdLS}
\end{table}

\subsection{\label{ssec:LSoptProps}Optical Properties}

\begin{figure}[h]
\centering
\includegraphics[width=0.65\textwidth]{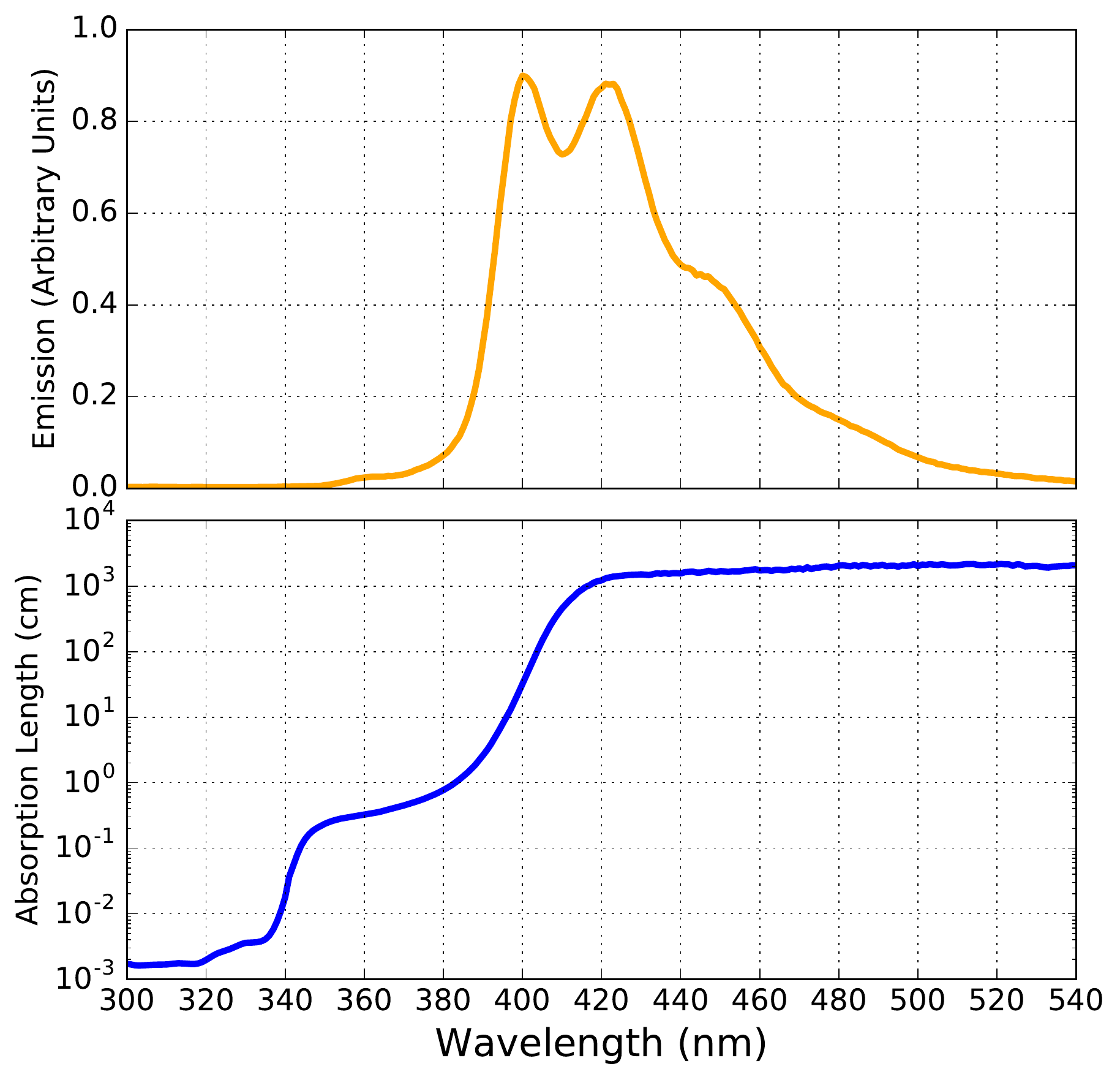}
\caption{\small Wavelength dependence of key optical properties of the GdLS. \textit{Top:} The emission spectrum of the GdLS scintillation light (in arbitrary units). \textit{Bottom:} The absorption length of GdLS. A full color version of this image is available online.\label{fig:lsOptical}}
\end{figure}

Energy deposited in the GdLS causes excitations of LAB molecules. These excitations are transferred to PPO~\cite{Furst:1954} resulting in the emission of fluorescent light, primarily below \SI{\lesssim380}{\nm}~\cite{Xiao:2010}. The PPO fluorescent light is absorbed and re-emitted at longer wavelength~\cite{Kallman:1951}
by bis-MSB (wavelength shifter) according to the spectrum shown at the top of Fig.~\ref{fig:lsOptical}. The emission spectrum peaks in the \SIrange{410}{425}{\nm} range~\cite{Beriguete:2014gua}, well matched to the photocathode response of typical PMTs.

The GdLS light yield in response to a \SI{662}{\keV} \Pgg is \SI{53}{\percent} of anthracene~\cite{Ding:2008zzb}. A similar electromagnetic energy deposited in anthracene needs \SI{55+-5}{\eV} per transmitted \SI{452}{\nm} photon~\cite{Wright:1955a,Wright:1955b}, resulting in \SI{10000+-1000}{photons\per\MeV} in GdLS. The description of light yield for different specific energy depositions $\mathrm{d}E/\mathrm{d}x$ will be described here by a generalized~\cite{Chou:1952} Birk's Law~\cite{Birks:1951}:
\begin{equation}
\label{equ:birk}
\frac{\mathrm{d}L}{\mathrm{d}x} = Y \frac{ \sfrac{\mathrm{d}E}{\mathrm{d}x} }{ 1 + kB \left(\sfrac{\mathrm{d}E}{\mathrm{d}x}\right) + C \left( \sfrac{\mathrm{d}E}{\mathrm{d}x}\right)^{2} }
\end{equation}
Where $Y$, the light yield, as well as $kB$ and $C$ are parameters
that are determined separately
for electron recoils in response
to \Pgg's and for \Pga's~\cite{Taylor:1951} by the calibrations described in Sec.~\ref{sec:detResponse}.

Optical photons traversing the GdLS may be absorbed by the liquid.  The $e$-folding absorption length is shown at the bottom of Fig.~\ref{fig:lsOptical}. A long absorption length is desired for increased detection efficiency. The GdLS absorption length rises from \SI{\approx1}{\meter} at \SI{410}{\nm} to \SI{>10}{\meter} at \SI{425}{\nm}, resulting in little self-absorption of the emitted scintillation light.

\subsection{\label{ssec:GdScreening}Screening of Gd Additives}

A \SI{0.307}{\kg} sample of the same chelated $\mathrm{Gd(TMHA)}_{\mathrm{3}}$ compound dissolved in the Screener GdLS was counted for \SI{18}{\day} in a HPGe detector in the Berkeley Low Background Facility in the Black Hills Underground Campus~\cite{Mount:2017iam}. Concentration causes the specific activity in \si{\mBqkg} as measured for the chelated compound to be 250 times higher than the specific activity for GdLS. Results from the HPGe exposure are reported in Tab.~\ref{tab:gdScreening}. The measured level of \IUtTeM may have been caused by radon contamination in the sample before the screening had started. The strong \Pgg-line from \IUtTF (the head of the \IUtTFE subchain) was not detected. Due to the small sample mass and short counting period, the sensitivities achieved in this assay were too low for adequate characterization of the Gd impurities. However, the detection of \ILuoSs was important for understanding the Screener data.

\begin{table}[h!]
\centering
\footnotesize
\caption{\small Results from HPGe counting of purified $\mathrm{Gd(TMHA)}_{3}$ powder given in \si{\mBqkg} of $\mathrm{Gd(TMHA)}_{\mathrm{3}}$ and in mBq/kg of GdLS, loaded to 0.1\% Gd by mass. To meet the LZ OD specifications, an average specific activity \SI{\leq0.07}{\mBqkg} of GdLS is needed. Limits are given at 68\% CL and a systematic uncertainty of 10\% is assumed on all values. \label{tab:gdScreening}}
\begin{tabular}{@{\extracolsep{0.01\textwidth}} l c c } \hline
{\textbf{Isotope or}} & \multicolumn{2}{c}{\textbf{Measured}} \\ 
 {\textbf{Subchain}} & \textbf{mBq/(kg $\textrm{Gd(TMHA)}_{3}$)} & \textbf{mBq/(kg GdLS)}  \\ \hline
\IUtTeE & $<259$ & $<1.04$ \\
\IUtTeM & $23\pm5$ & $0.092\pm0.02$ \\
\IUtTFE & $<2.8$ & $<0.011$ \\
\IUtTFL & $26\pm10$ & $0.10\pm0.04$ \\
\IThtTtE & $<6.7$ & $<0.027$ \\
\IThtTtL & $<5.1$ & $<0.020$ \\
\IKfz & $<56$ & $<0.22$ \\
\ILaoTe & $<1.4$ & $<0.0055$ \\
\ILuoSs & $75\pm18$ & $0.30\pm0.07$ \\
\hline
\end{tabular}
\end{table}

\section{\label{sec:DataCollection}Data Collection and Calibrations}

The Screener was shipped empty to the Davis Laboratory, where it was filled and commissioned in a cleanroom. When work was not being performed, the detector was kept in a dark box continuously purged with nitrogen gas in order to limit exposure to the \SI{\approx300}{\BqmT}~\cite{Heise:2017rpu} radon level in the Davis Laboratory air. 

After commissioning, the detector was rolled out of the cleanroom on a cart and lifted off its stand with an overhead hoist. The detector was then lowered into the water tank and suspended by polyethylene ropes from hooks on the water tank ceiling as shown in Fig.~\ref{fig:screener_in_tank}. Cables and other services, including a gas line, an optical fiber, and radioactive sources were fed through a port in the water tank top. A check for light leaks around the water tank port was performed once the PMTs were biased.

Data were taken in two runs. ``Run 1" data were taken between November 2016 and early January 2017, using $23.7\pm0.1$~kg of GdLS. Data in ``Run 2" were taken between mid-January 2017 and February 2017, using the same scintillator mixture, but without Gd loading. In Run 2 the LS mass was $23.2\pm0.1$~kg.

\subsection{\label{ssec:PMTcal}PMT Calibrations}

The response of the PMTs was monitored $\sim$weekly with two types of single photoelectron calibrations. Light from a \SI{420}{\nm} LED in the DAQ system was fed through an optical fiber through the water tank port to the Screener to provide dedicated single photoelectron samples. Sometimes these calibration runs indicated drift in the PMT or amplifier response, so single photoelectron pulses in the midst of data taking  were identified offline and used to correct drift during running. The PMT bias voltages were sometimes adjusted to equalize the single photoelectron response and were typically \SI{1385}{\volt}, \SI{1282}{\volt}, and \SI{1377}{\volt}.

\subsection{\label{ssec:Sourcecal}Radioactive Source Calibrations}

Several radioactive \Pgg sources were used to calibrate the Screener response  \emph{in-situ} to energy deposited by electron recoils. For these calibrations the Screener was hoisted up to the water tank top, the sources were attached, and the assembly was lowered back into the tank. \ICsoTS and \IThtte \Pgg-ray disk sources, in a tungsten collimator with a \SI{1.75}{\cm}-diameter aperture, were deployed on top of the Screener, and provide the principal electron recoil calibrations.

The detector response to energy deposits from \Pga's, \Pgb's, and \Pgg's was calibrated with a thoron (\IRnttz) source and its decay products (Fig.~\ref{fig:ThtTtsub}). The thoron source consisted of nitrogen gas flowed through a \IThtte source; the gas was then bubbled into the the Screener scintillator volume. The signals from the thoron source have a distinct time dependence which we exploit: while nitrogen gas is flowing, a merged peak from the \Pga-decays of \IRnttz and \IPotos appear due to their relatively short half-lives (\SI{55}{\second} and \SI{0.14}{\second}, respectively). These peaks promptly disappear once nitrogen flow is stopped. The population of daughter \IPbtot atoms then feeds the rest of the chain according to its \SI{10.6}{\hour} half-life.  After roughly four days the thoron decay products had decayed to a level below that of the background rate.

Two thoron calibrations were performed in Run 1: the first after initial deployment of the detector in the water tank and the second prior to its removal. In Run 2, a single thoron calibration was performed prior to detector removal.

\section{\label{sec:detResponse}Detector Response Model}

A detailed model of the detector geometry is implemented in the \textsc{Geant4} toolkit. The detector simulation models the deposition of energy in the detector materials and the generation, propagation, and detection of optical photons born in the LS.

\subsection{Optical Model}

The optical model includes the wavelength-dependent emission and absorption properties of the LS discussed in Sec.~\ref{ssec:LSoptProps}. The acrylic absorption length and the indices of refraction for GdLS and acrylic are also included~\cite{Littlejohn:2012qpa}.

\begin{figure}[h]
\centering
\includegraphics[width=0.65\textwidth]{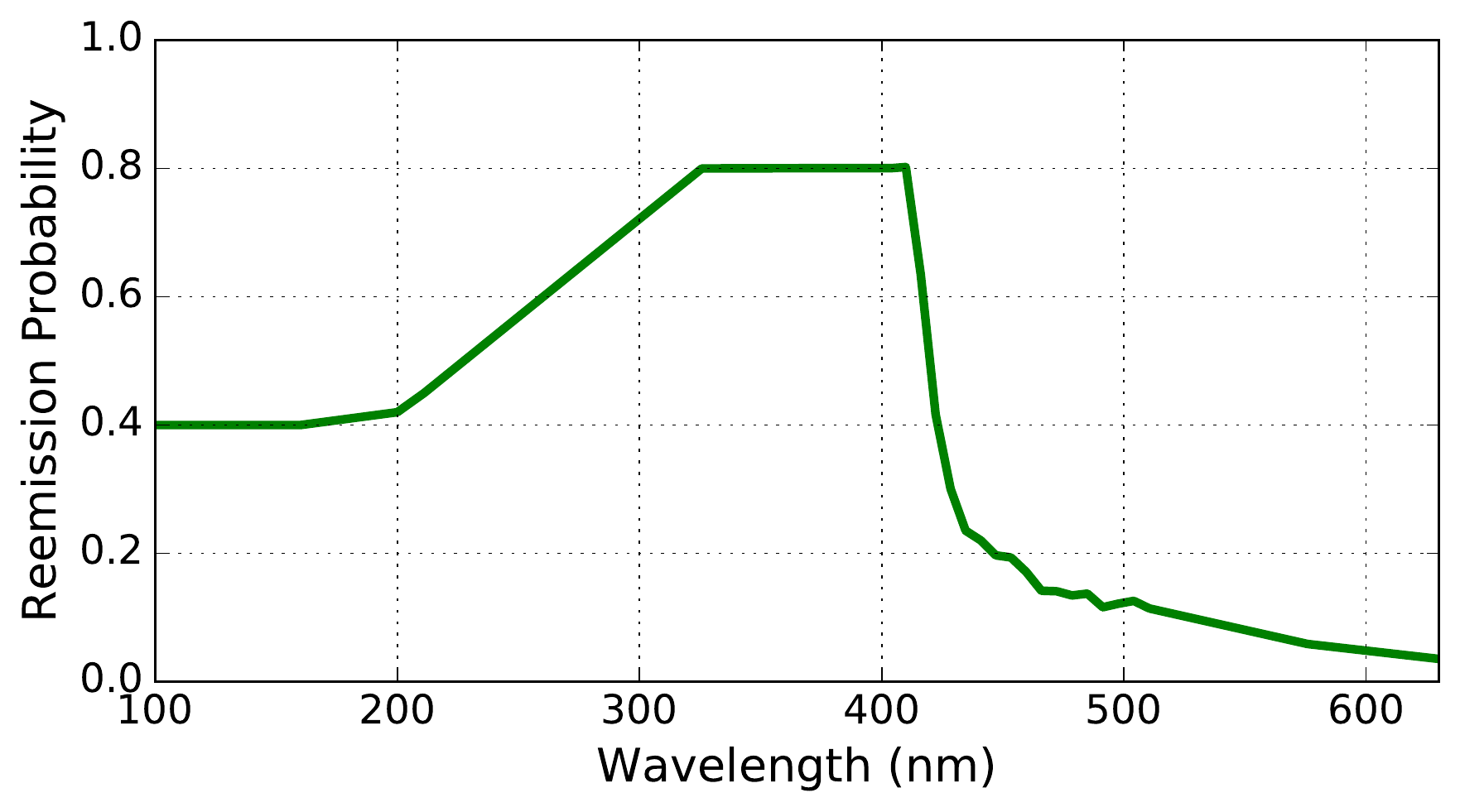}
\caption{\small GdLS re-emission probability implemented in the optical model. A full color version of this image is available online.\label{fig:lsReemission}}
\end{figure}

The model also includes the re-emission probability of the GdLS shown in Fig.~\ref{fig:lsReemission}. This is the probability that an absorbed photon of a given energy is re-emitted according to the bis-MSB spectrum in Fig.~\ref{fig:lsOptical}. We found that agreement between \IThtte \Pgg calibration data and the simulation could only be obtained by including this effect in the optical response of the GdLS. The \SI{2.6}{\MeV} \Pgg from this source produces Compton electrons above the \SI{180}{\keV} Cherenkov threshold in GdLS. With the addition of the re-emission process in the model, the far-UV Cherenkov photons that are more heavily absorbed in GdLS can be re-emitted at longer wavelengths and detected more efficiently.

\begin{figure}[h!]
\centering
\includegraphics[width=0.65\textwidth]{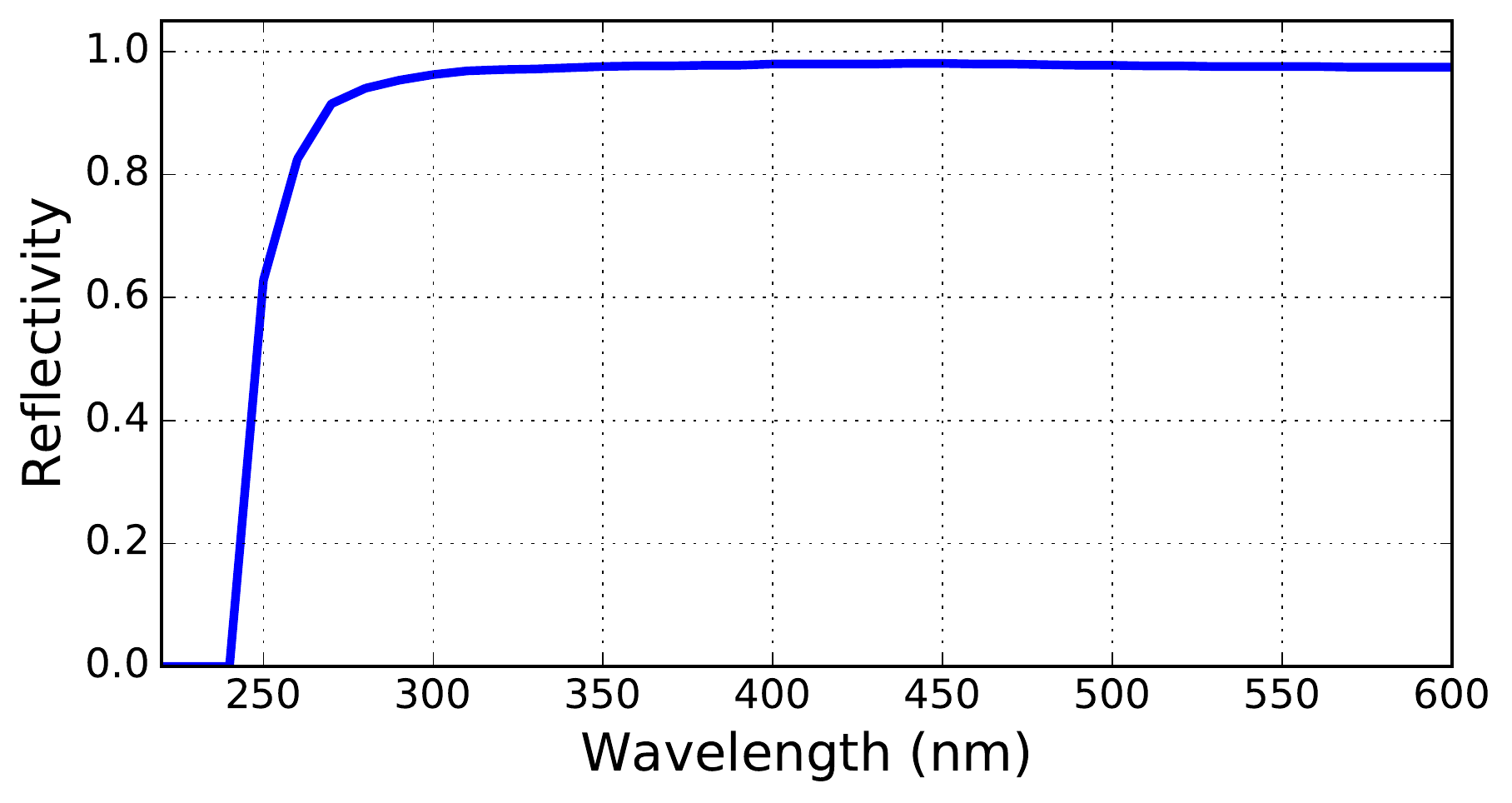}
\caption{\small Wavelength dependence of the 1085D Tyvek reflectivity used in the detector optical model. A full color version of this image is available online.\label{fig:tyvek}}
\end{figure}

The Tyvek surrounding the detector vessel is assumed to have the diffuse reflectivity dependence shown in Fig.~\ref{fig:tyvek}. The shape of this dependence was measured in Ref.~\cite{Janecek:2012}, while the overall scale was determined during surface commissioning of the detector where scintillator paddles above and below the vessel were used to tag Cherenkov events caused by muons traversing the acrylic vessel. The maximum reflectivity in the model is \SI{98.1}{\percent}. The reflectivity and the LS light yield parameters are somewhat degenerate in the task of modeling the total number of photons collected by the PMTs. An overestimate of the true Tyvek reflectivity will result in a corresponding underestimate of the scintillator light yield, but the total number of collected quanta will largely remain the same.

The generation of optical photons from energy deposits in the LS is implemented using the model in Eq.~(\ref{equ:birk}). The parameters $Y$, $kB$, and $C$ are determined by tuning the simulation to the calibration data collected in each run.

\subsection{Calibration of Simulation Response\label{ssec:simTuning}}

\begin{figure}[b!]
\centering
\begin{subfigure}{0.49\textwidth}
\includegraphics[width=\textwidth]{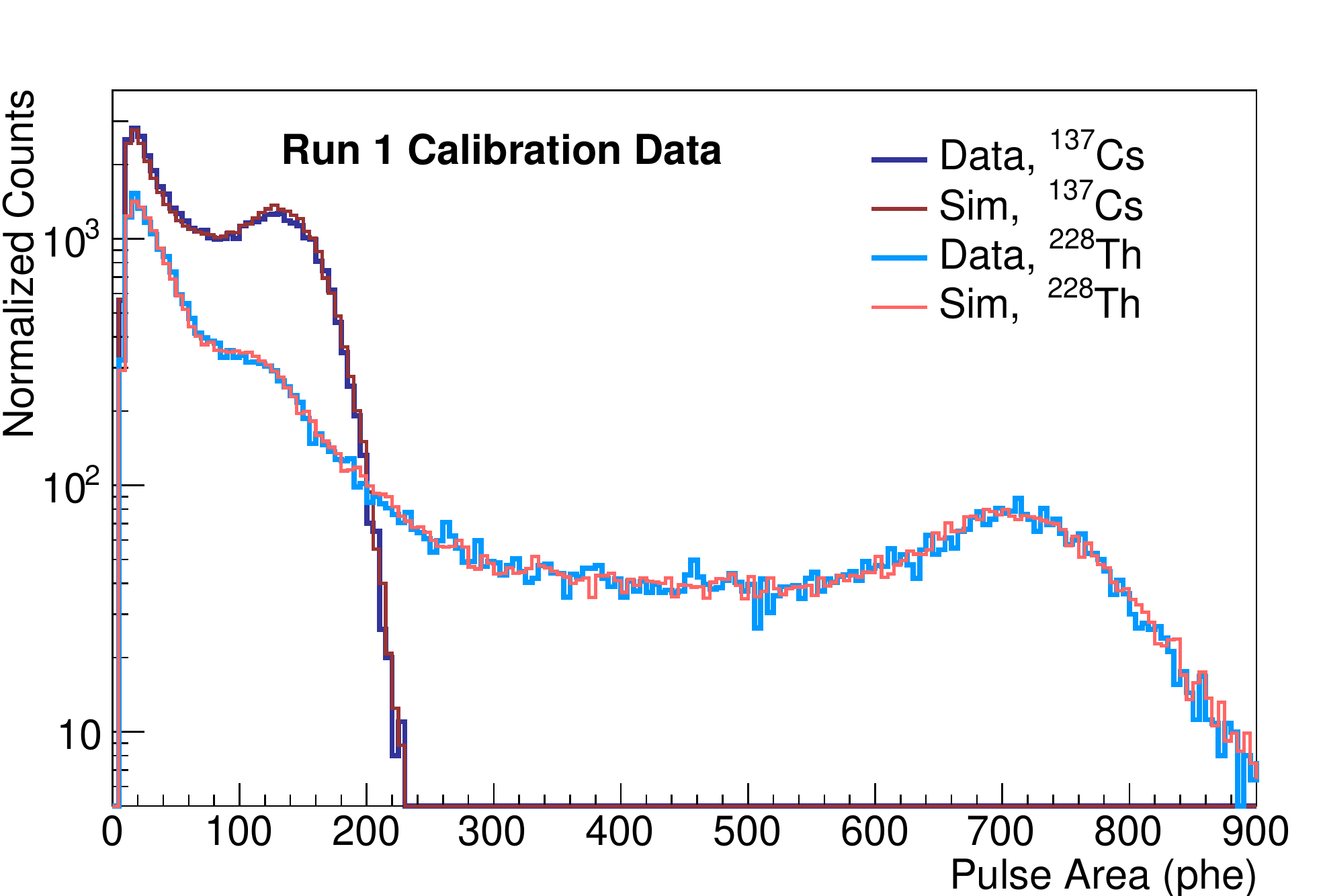}
\end{subfigure}
\begin{subfigure}{0.49\textwidth}
\includegraphics[width=\textwidth]{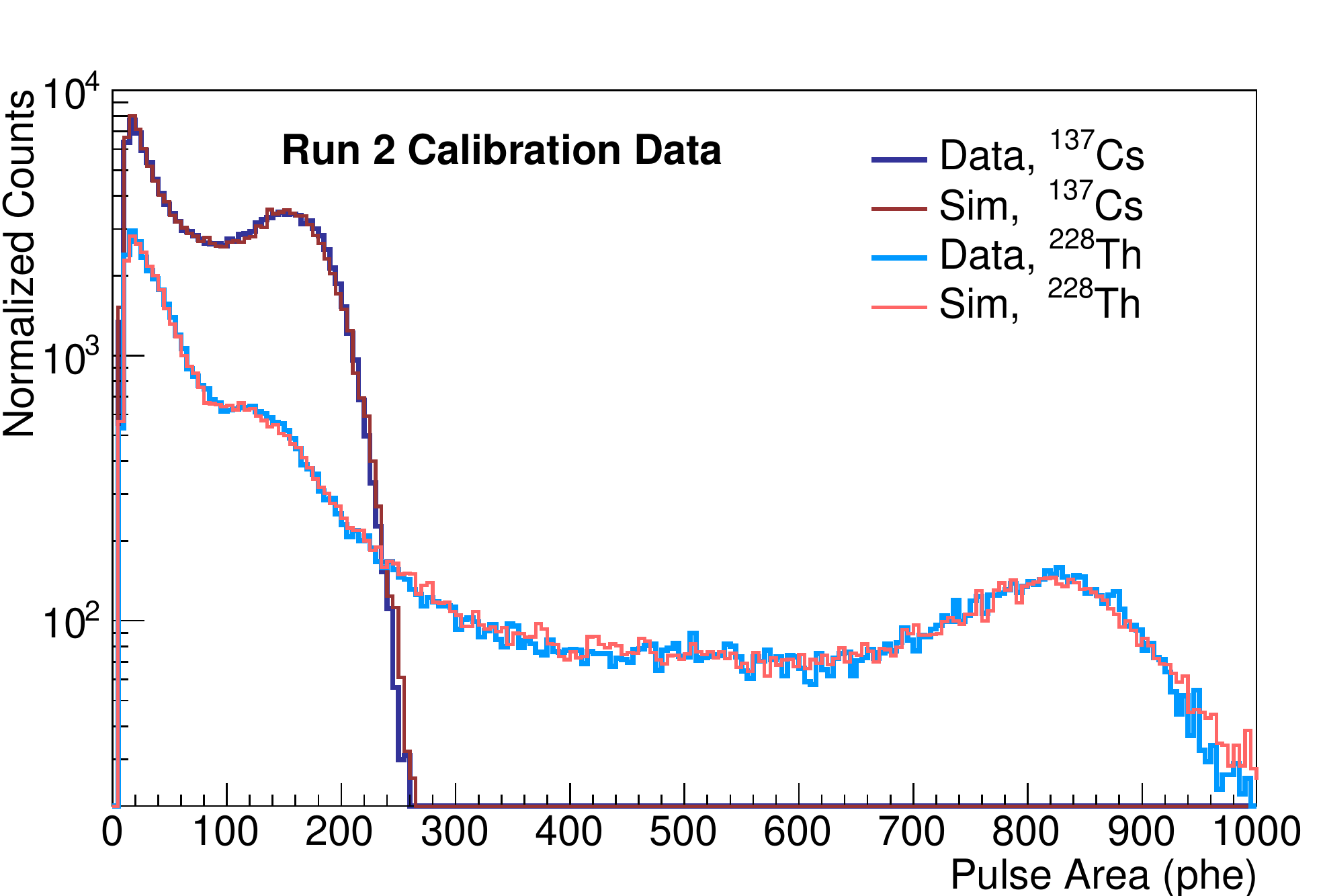}
\end{subfigure}
\caption{\small Pulse area spectra collected during calibrations with the \ICsoTS and \IThtte sources. Overlaid are the simulated spectra obtained using the best fit $Y$ and $kB$ parameters. A full color version of this image is available online.\label{fig:gammaCal}}
\end{figure}

To calibrate the simulated detector response to energy deposits from electrons, different values of $Y$ and $kB$ are simulated and compared with \ICsoTS and \IThtte \Pgg calibration data until good agreement via a $\chi^{2}$ fit is obtained. The parameter $C$ is set to zero for electron energy deposits. The $kB$ parameter accounts for non-linearity in the scintillator response. Pulse area spectra for the best-fit parameters are shown with data in Fig.~\ref{fig:gammaCal}. The spectra of true energy deposits in the LS from these simulations show that the peaks in data are formed by the merging of a Compton edge and the corresponding photopeak.

The yield in phe/MeV in each run is determined by simulating energy deposits from electrons uniformly throughout the LS volume. The resulting distributions are skew Gaussian, a consequence of variation in light collection along the LS chamber vertical axis. The size of this variation is \SI{\approx20}{\percent} across the full height. Simulations of uniform \SI{1}{\MeV} electrons in the LS give corresponding photoelectron yields of $321\pm6$~phe/MeV in Run 1 and $380\pm12$~phe/MeV in Run 2 using the mean values of the resulting skew Gaussian distributions. The error bars are obtained by varying $Y$ and $kB$ within their uncertainties and repeating the uniform electron simulations. These factors form the ``approximate energy" scale used throughout this work; the non-linearity of the scintillator response with energy making it approximate. For the \Pga fit results shown in Sec.~\ref{ssec:results}, an ``approximate quenched energy'' scale is shown, which reports \Pga energies as their electron-equivalent energy.

\begin{figure}[h!]
\centering
\begin{subfigure}{0.49\textwidth}
\includegraphics[width=\textwidth]{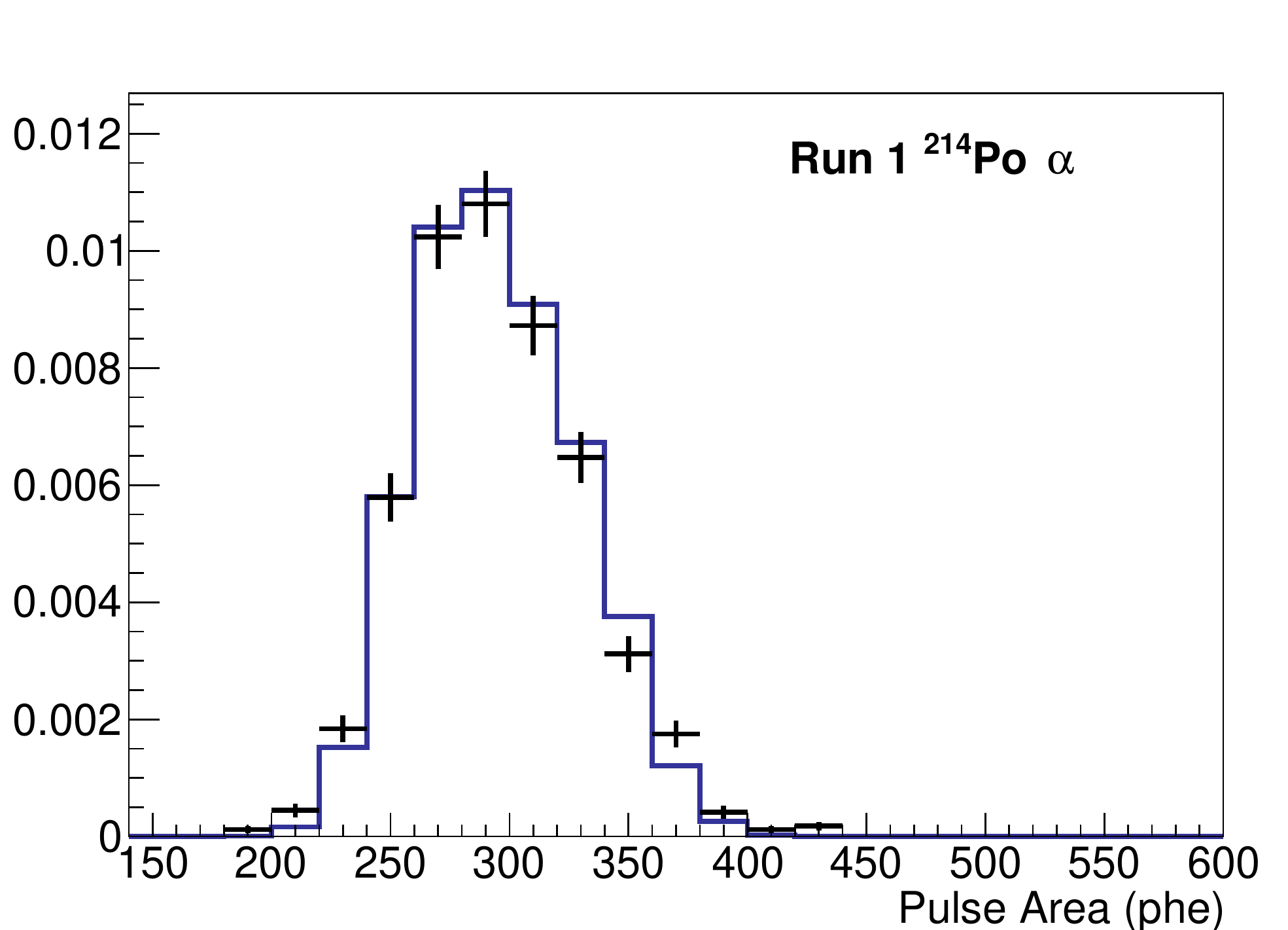}
\end{subfigure}
\begin{subfigure}{0.49\textwidth}
\includegraphics[width=\textwidth]{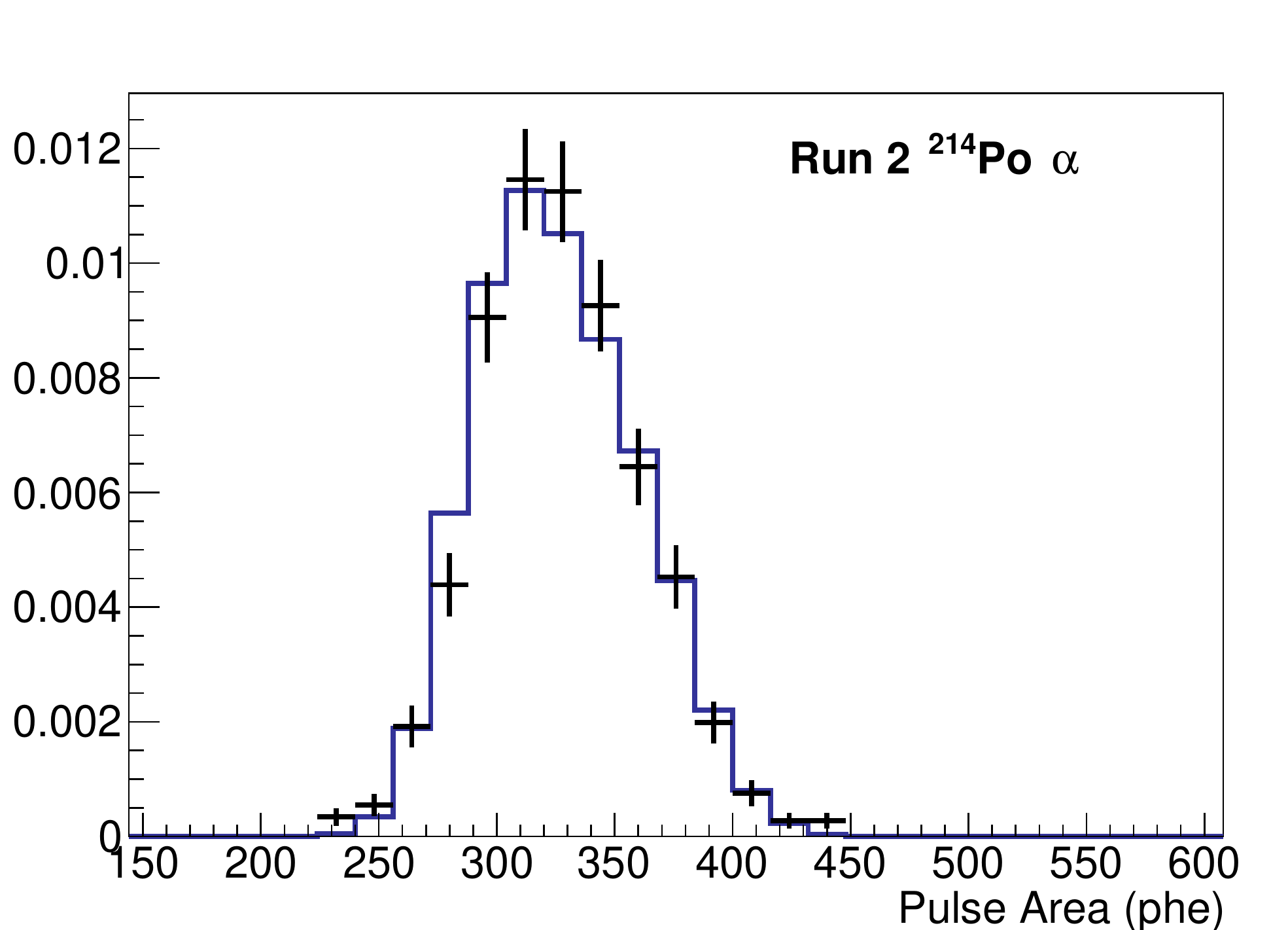}
\end{subfigure}
\caption{\small Pulse area spectra of \IPotof \Pga-decays in both runs obtained from BiPo coincidence cuts. The simulated spectra obtained using the best fit $kB$ and $C$ parameters are shown in blue. Both spectra have unit normalization. A full color version of this image is available online.\label{fig:alphaCal}}
\end{figure}

The detector response to \Pga events is similarly determined by simulating a set of \Pga energies for different values of $kB$ and $C$. The mean number of collected phe in the simulation is compared to that from each peak in data with a $\chi^{2}$.

Early data from both runs provide an \Pga peak from dissolved \IRnttt. The same data is analyzed for \Pgb-\Pga coincidences from $^{214}\textrm{Bi}$ to $^{214}\textrm{Po}$ (``BiPo") decays (see Section \ref{ssec:bipo}), allowing the construction of an \Pga peak from $^{214}\textrm{Po}$. This peak in data is shown in Fig.~\ref{fig:alphaCal} along with the simulated response using the best fit $kB$ and $C$ parameters. Calibration data from the flow-through thoron source provides a merged peak from $^{220}\textrm{Rn}$ and $^{216}\textrm{Po}$ while the gas is flowing. Data collected after the flow has stopped provides peaks from $^{212}\textrm{Bi}$ and $^{212}\textrm{Po}$, the former through its direct \Pga-decay (branching ratio \SI{36}{\percent}) and the latter by identifying the prompt $^{212}\textrm{Po}$ \Pga's following the \Pgb-decay of $^{212}\textrm{Bi}$ (branching ratio \SI{64}{\percent}). Finally, a merged, low energy peak from \IGdoFt/\ISmofS decays is present in Run 1 data around \SI{\approx30}{phe} where GdLS was used (Fig.~\ref{fig:bkgSpectra} left).

Comparisons of the same \Pga peak distributions during different thoron calibration periods show that the phe yield had changed during the calibration. It is likely that oxygen dissolved in the LS during the filling process was stripped by the active bubbling of nitrogen gas during the calibration. As a result, not all the available \Pga peaks are used for determining the $kB$ and $C$ parameters. In Run 1, the \Pga peaks from \IGdoFt/\ISmofS, \IRnttz/\IPotos, \IBitot, \IPotof, and \IPotot are used, while in Run 2 the peaks from \IRnttt, \IRnttz/\IPotos, and \IPotof are used. Between calibrations, the phe yield is monitored using the position of the \IPotof peak and found to be stable to within \SI{12}{\percent} in Run 1 and \SI{6}{\percent} in Run 2. 

\begin{table}[h]
\centering
\footnotesize
\caption{\small Scintillator parameters deduced from tuning the \textsc{Geant4} detector simulation to \Pgg and \Pga calibration data. Also shown are the associated $\chi^2/\textrm{ndf}$ values for each of the fits. \label{tab:simValues}}
\begin{tabular}{l c c} \hline
\textbf{Parameter} & \textbf{Run 1} & \textbf{Run 2} \\ \hline
LS Light Yield, $Y$ ($\textrm{photons}~\textrm{MeV}^{-1}$) & $8783 \pm 57$ & $10202 \pm 138$ \\
$kB$ \Pgg's ($\textrm{cm}~\textrm{MeV}^{-1}$) & $0.0299 \pm 0.0016$ & $0.0288 \pm 0.0036$ \\
$kB$ \Pga's ($\textrm{cm}~\textrm{MeV}^{-1}$) & $0.00512\pm0.00041$ & $0.00744\pm0.00070$ \\
$C$ \Pga's ($\textrm{cm}^{2}~\textrm{MeV}^{-2}$) & $(2.43\pm0.43)\times10^{-6}$ & $(0.68\pm0.54)\times10^{-6}$ \\
\hline
\hline
\Pgg Fit $\chi_{\textrm{min}}^2/\textrm{ndf}$ & 55.7/46 & 56.7/46 \\
\Pga Fit $\chi_{\textrm{min}}^2/\textrm{ndf}$ & 3.3/3 & 1.5/1 \\
\hline
\end{tabular}
\end{table}

Tab.~\ref{tab:simValues} summarizes the best fit simulation parameters used in each run. The lower light yield (and hence the lower phe yield) observed in Run 1 is thought to be from exposure of the LS to air during the filling process. This is also believed to be the cause of the higher \IRnttt rate at the beginning of that run.

The parameters $kB$ and $C$ describing \Pga quenching in the LS are different for the two runs. The difference arises from the inclusion of the low energy \IGdoFt/\ISmofS \Pga peak in the fit for Run 1. The values found here are generally consistent with those found in~\cite{VonKrosigk:2015yio} where it is shown that a range of $kB$ and $C$ values can result in similar overall quenching. The differing values are therefore not a concern, as our focus is on the response to \Pga's of energy \SIrange{4}{9}{\MeV}. This range contains \Pga-decays from the uranium and thorium decay chains.

\subsection{\label{ssec:psd}Pulse Shape Discrimination}

The timing profile of emitted photons in the LS is dependent on the particle type. Energy deposits from particles such as \Pga's that produce large amounts of ionization along their track produce photons with a characteristically slower time profile compared to electrons~\cite{OKeeffe:2011dex}. This allows for PSD between \Pga and \Pgg/\Pgb events.

\begin{figure}[h]
\centering
\includegraphics[width=0.98\textwidth]{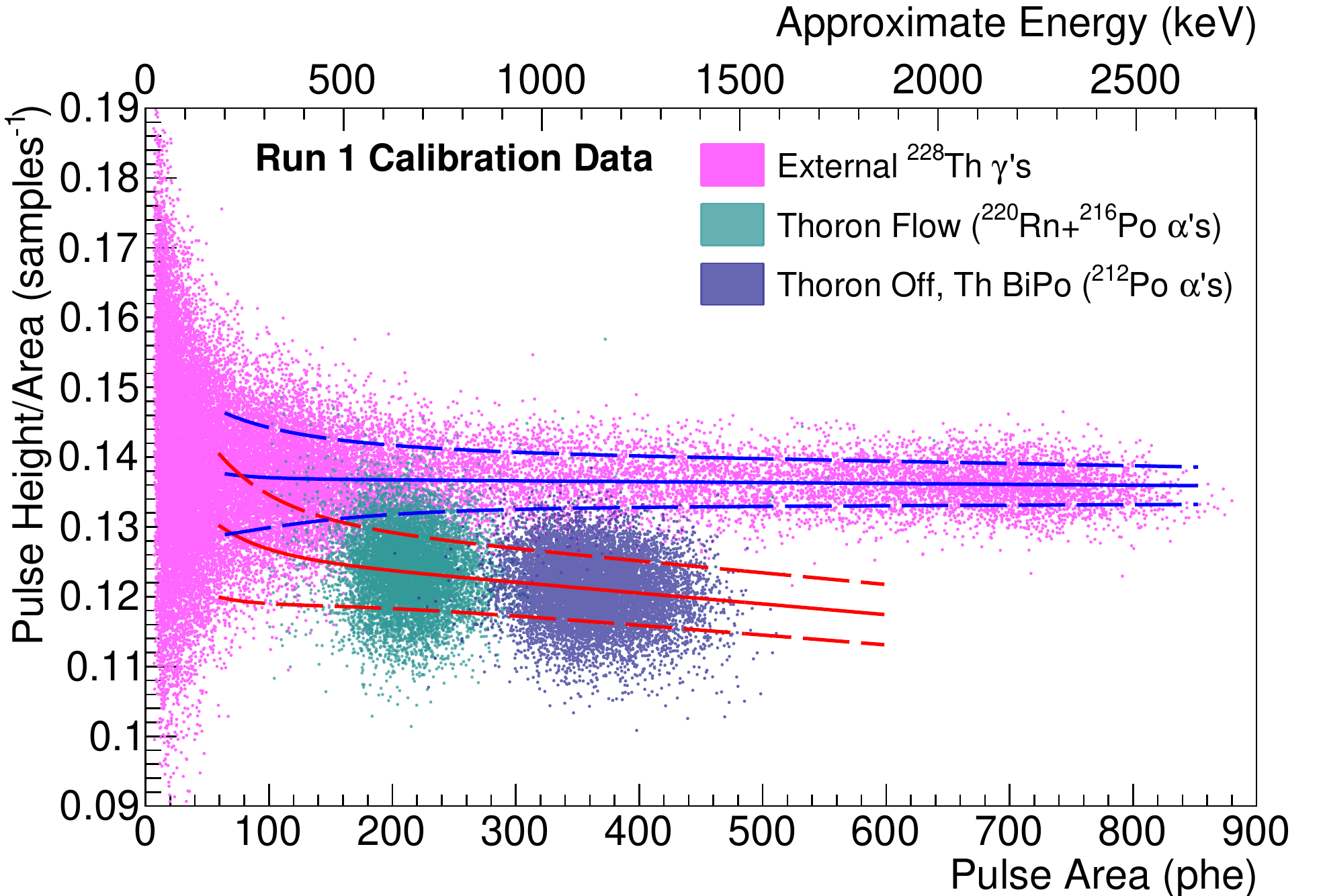}
\caption{\small Calibration data with \Pga's (cyan and blue) and \Pgg's (magenta) shown in PSD space. Overlaid are the fit \Pgg (blue) and \Pga (red) band means (solid) and $\pm 1\sigma$ (dashed) curves. A full color version of this image is available online.\label{fig:calPSD}}
\end{figure}

We use a simple parameter, the ratio of the pulse height to area as the discrimination variable. For a fixed pulse area, the characteristically slower pulses from \Pga's have a smaller height than those from electrons, providing separation between pulses from these particle types. In Fig.~\ref{fig:calPSD}, pulses from the external \IThtte source (magenta) are shown with data from the flow-through \IRnttz source (cyan and blue) in the height-to-area ratio versus area plane. The separation between \Pga events and electron events is clear. As demonstrated in~\cite{OKeeffe:2011dex}, better separation is possible when oxygen is removed from the LS, e.g. by thoroughly bubbling with nitrogen gas.

The \Pgg and \Pga calibration data in each run are profiled by fitting a Gaussian function to the height-to-area ratio distributions found in slices of pulse area. The means and variances obtained from the fit in each slice are subsequently fit to a linear plus exponential model, thus defining \Pgg/\Pgb and \Pga bands within the PSD space. The bands are shown also in Fig.~\ref{fig:calPSD}.

Selections of \Pga and \Pgg/\Pgb events are made using cuts on the PSD parameter. For \Pga events we take the PSD parameter to be $3\sigma$ below the calibrated \Pgg/\Pgb band mean. For \Pgg and \Pgb events we define a cut on the PSD parameter which takes the linear plus exponential form used to describe the band shape. The cut is at least $2\sigma$ above the calibrated \Pga band mean at all energies.

The leakage of \Pgg/\Pgb events into the \Pga-selected data is negligible. In Run 1 the expected leakage is 26 events, while a total of 5823 events are accepted by the cut. In Run 2, the expected leakage is 9 events with 911 events passing the selection. The expected leakage of \Pga events into the \Pgg/\Pgb-selected data is 65 and 16 events in Run 1 and Run 2, respectively, while the number of accepted \Pgg/\Pgb events in the two runs are 7521 and 4702.

\begin{figure}[h]
\centering
\begin{subfigure}{0.49\textwidth}
\includegraphics[width=\textwidth]{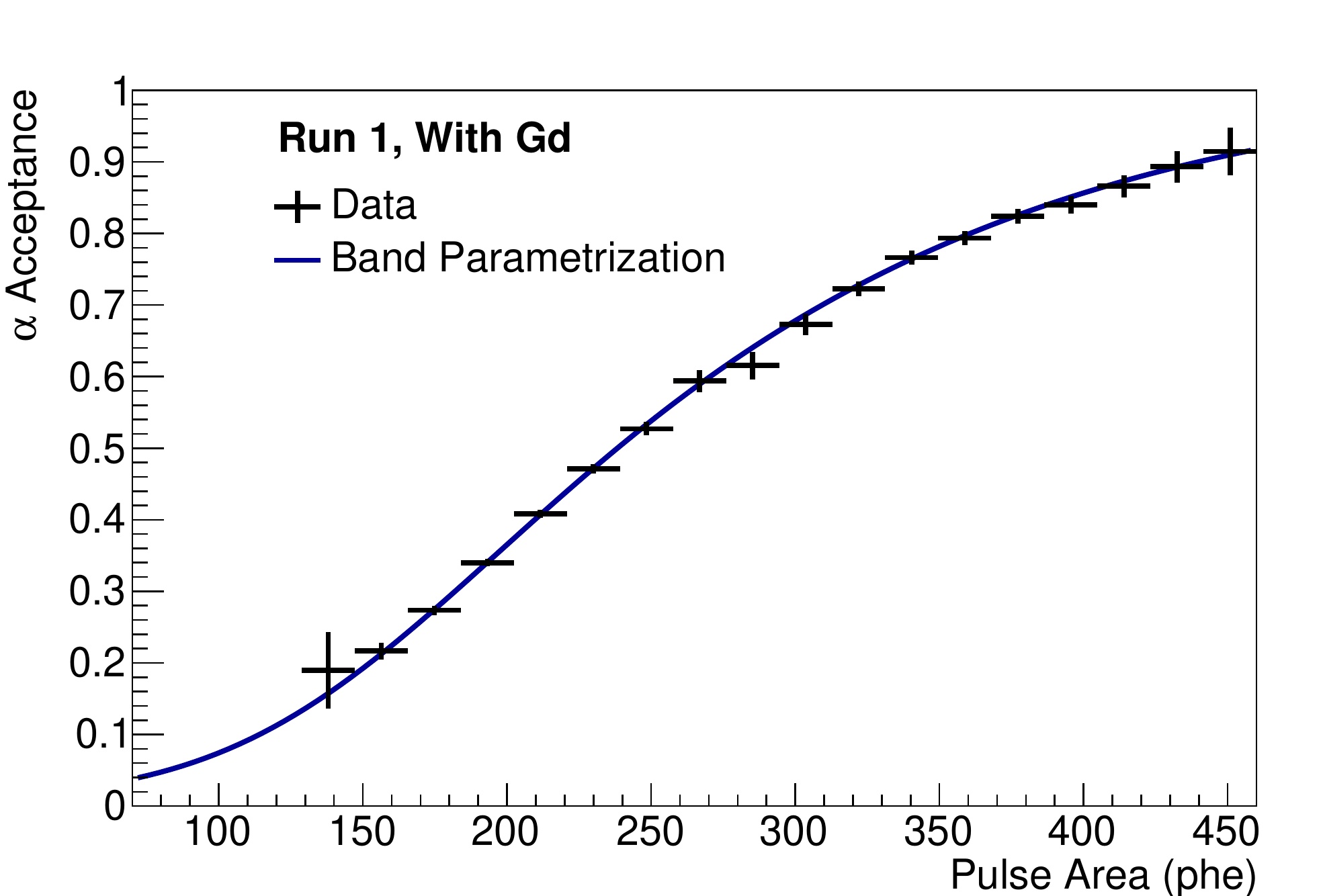}
\subcaption{\label{subfig:alphaeff}}
\end{subfigure}
\begin{subfigure}{0.49\textwidth}
\includegraphics[width=\textwidth]{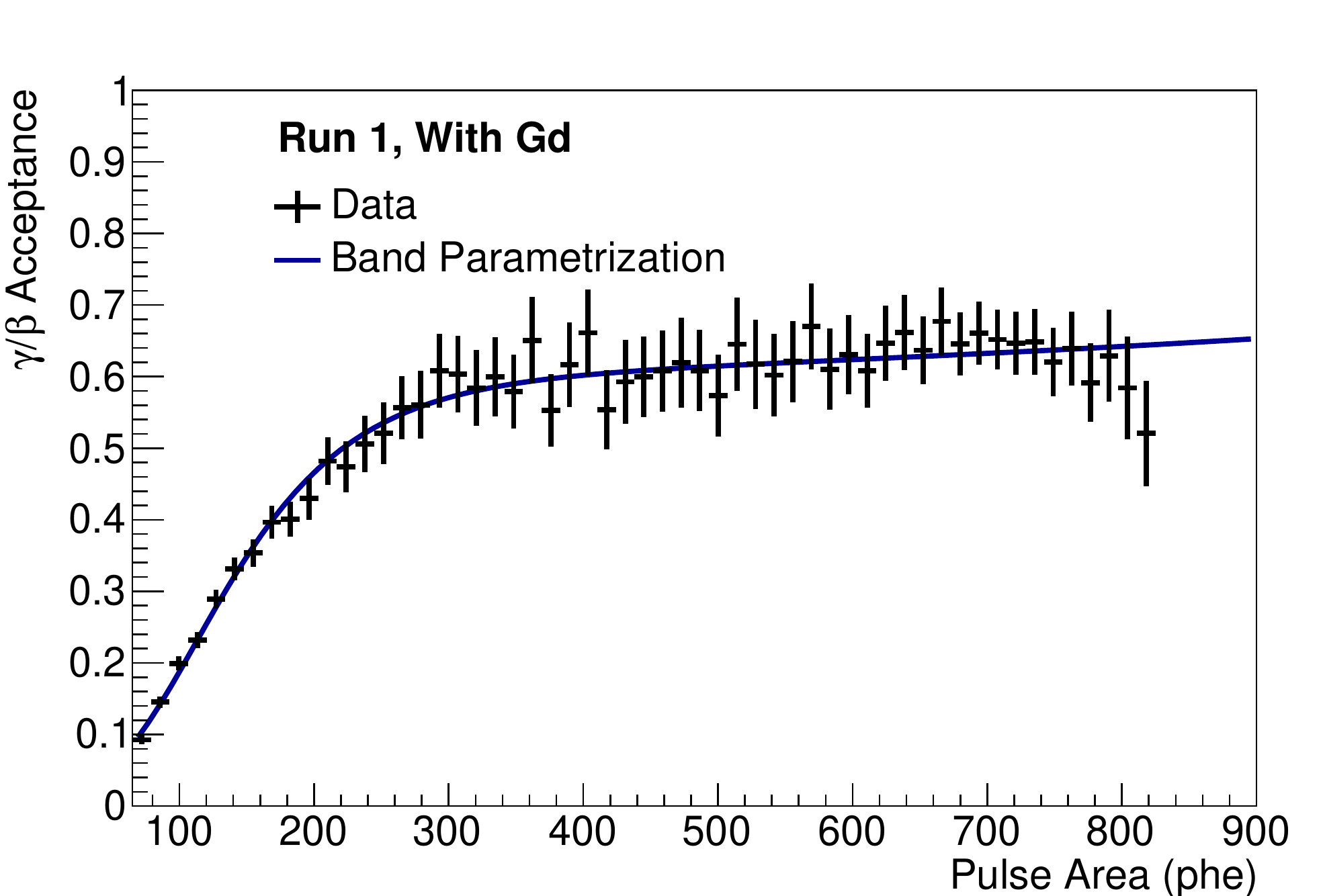}
\subcaption{\label{subfig:gammaeff}}
\end{subfigure}
\caption{\small (a) Acceptance of \Pga events and (b) \Pgg/\Pgb events as a function of pulse area computed using the Gaussian band parameterization of calibration data (blue) plotted with that computed using the data itself (black). Note that these are not fits. A full color version of this image is available online.\label{fig:eff}}
\end{figure}

The cut efficiencies in Run 1 are shown in Fig.~\ref{fig:eff}. The efficiencies for both cuts are calculated both directly from thoron and \Pgg source calibration data and using the parameterization of the bands. In the region where calibration data is available, good agreement is found between the band parameterization and the data, justifying the use of the band model in regions where there is no calibration data.

\FloatBarrier

\section{\label{sec:lowBkgData}Measurement of Backgrounds in the LS}

At the beginning of Run 1, the raw trigger rate was \SI{2.3}{\Hz}. Pulse area spectra from this early data show a clear peak from the \Pga-decay of \IRnttt that decays in time consistent with the expected half-life of \SI{3.8}{\day}. This source of \IRnttt is likely to have been dissolved in the scintillator while filling the vessel in the radon-rich underground environment. After approximately 21~days the trigger rate had plateaued to \SI{0.3}{\Hz}.

Similarly, spectra taken at the outset of Run 2 show a peak from \IRnttt. The trigger rate was \SI{0.32}{\Hz}. After 14~days the plateaued trigger rate was \SI{0.16}{\Hz}. The smaller contamination from radon in Run 2 was a result of an improved LS filling procedure.

The plateaued rate at the end of each run arises from sources of radioactivity in the detector construction materials, the water tank/cavern environment, and the radioimpurities in the scintillator. We refer to data collected at the end of each run as ``low-background" data. 

\begin{figure}[h!]
\centering
\begin{subfigure}{0.49\textwidth}
\includegraphics[width=\textwidth]{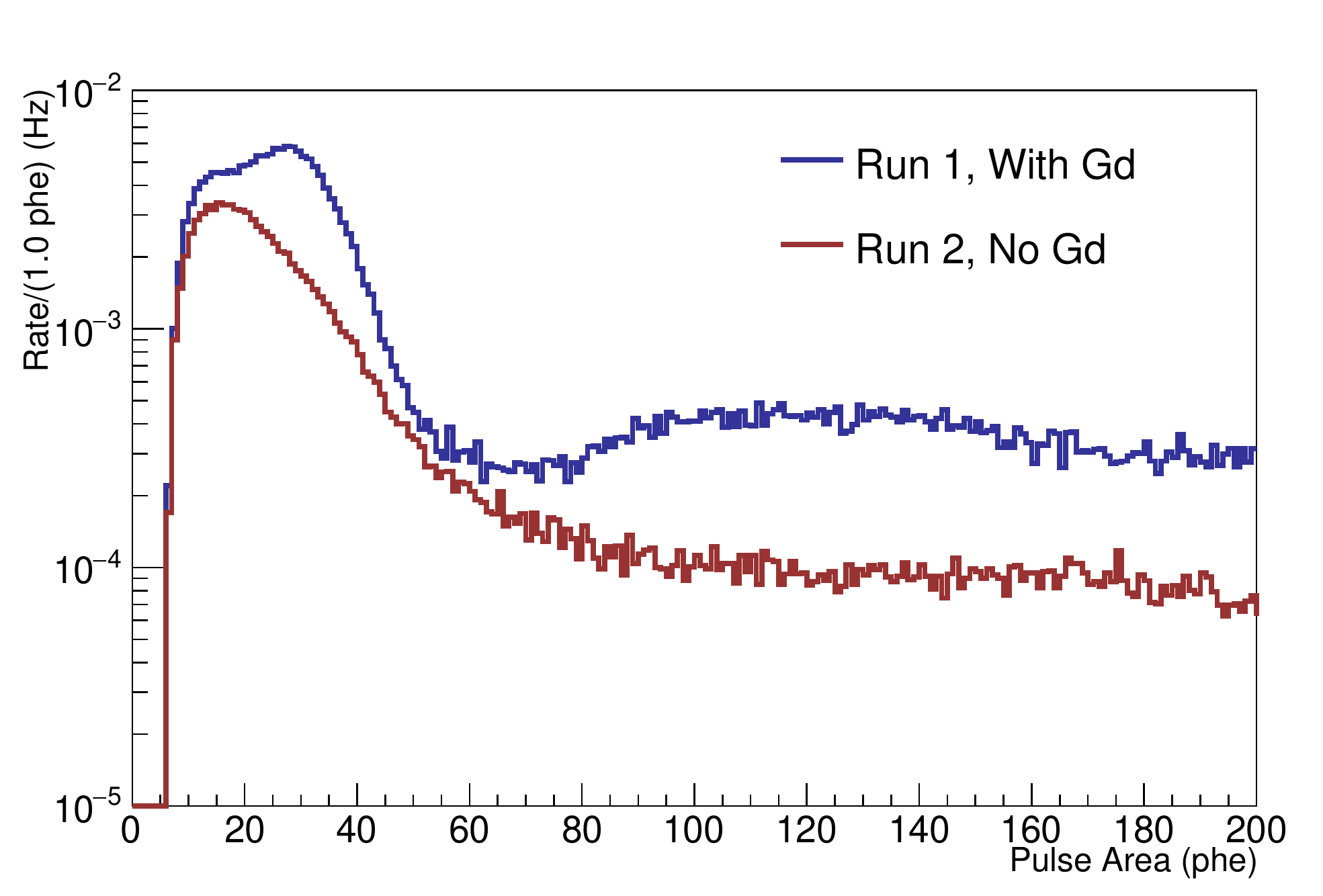}
\end{subfigure}
\begin{subfigure}{0.49\textwidth}
\includegraphics[width=\textwidth]{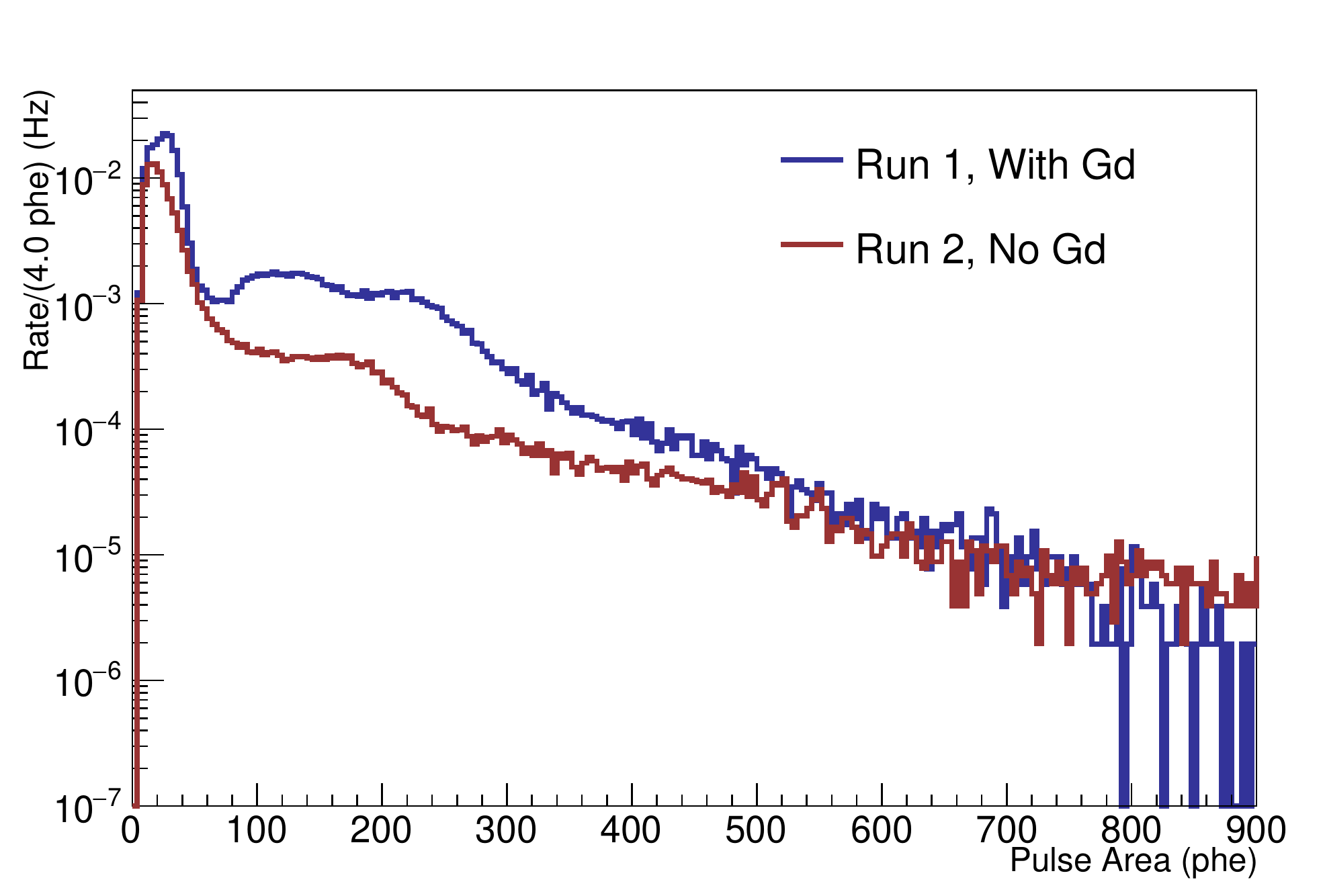}
\end{subfigure}
\caption{\small Run 1 and Run 2 low-background spectra shown in the low pulse area region (left) and over the full range of collected phe (right). The low energy peak at \SI{\approx30}{phe} from $^{152}\textrm{Gd}$/$^{147}\textrm{Sm}$ decays is visible in the Run 1 data on the left. Note that Run 1 and Run 2 do not share the same energy scale (phe/MeV). A full color version of this image is available online.\label{fig:bkgSpectra}}
\end{figure}

The inclusive low-background data counting rates above \SI{200}{\keV} in the GdLS and LS runs are \SI{78.6+-0.4}{\mHz} and \SI{20.0+-0.1}{\mHz}, respectively. The collected pulse area distributions for each run are shown in Fig.~\ref{fig:bkgSpectra}. The livetime durations are \SI{5.97}{\day} for Run 1 and \SI{11.83}{\day} for Run 2.

The low-background data are divided into two primary regions:

\begin{enumerate}
\item Low pulse area region, \SI{\lesssim 250}{\keV}. The dominant source of rate in this region results from the \Pgb-decay of \ICof. In Run 1, a peak results from the \Pga-decays of \IGdoFt and \ISmofS
that are quenched into this region.

\item High pulse area region, \SI{\gtrsim 250}{\keV}. Pulses in this region result from \Pgg backgrounds external to the LS and from the decays of the internal radioimpurities dissolved in the LS. PSD is used in this region to select \Pga and \Pgg/\Pgb events.
\end{enumerate}

Specifically, the high pulse area region is defined by pulse areas \SI{>65}{phe}, corresponding to \SI{>222}{\keV}, in Run 1 and \SI{>100}{phe}, corresponding to \SI{>281}{\keV}, in Run 2. The inclusive count rates measured with these thresholds are \SI{75.9+-0.4}{\mHz} and \SI{16.0+-0.1}{\mHz}, respectively.

\subsection{Measurement of \ICof Concentration \label{ssec:carbon14}}

A measurement of the ultra-low \ICof/\ICot ratio in pseudocumene-based LS was performed by the Borexino collaboration and found to be at the $10^{-18}$ level \cite{Alimonti:1998rc}. Here we follow nearly the same procedure to determine the \ICof concentration in our LS by fitting the low pulse area region of Run 2 with the \ICof \Pgb-decay ($Q=156$~keV) shape. 

The distribution of electron energies resulting from \Pgb-decay follows the form
\begin{equation}
\label{eqn:betashape}
N(W)\mathrm{d}W = pW(W_{0}-W)^{2}F(Z,W)C(W)\mathrm{d}W
\end{equation}
where $p$, $W$, and $W_{0}$ are the electron momentum, energy, and endpoint energy in units of the electron mass, respectively. The Fermi function, $F(Z,W)$, corrects the kinematic shape by accounting for Coulomb interactions between the daughter nucleus and the electron. The distribution is further altered by a shape factor $C(W)$, which we take to have the form:
\begin{equation}
C(W) = 1 + a W.
\end{equation}

As in~\cite{Alimonti:1998rc}, the tabulated values of the Fermi function in~\cite{springerBetaTable} were parameterized and used here.

The data in the low pulse area region is fit using the following convolution model:
\begin{equation}
\begin{aligned}
R(W)=N_{\textrm{C}} \int N(W') G(W',W)\mathop{}\!\mathrm{d}W' + N_{\textrm{bkg}}B(W)
\end{aligned}
\end{equation}

Here, $G(W',W)$ is a detector resolution function, which smears the true \Pgb-decay shape, $N(W')$. The \ICof and background shapes are weighted by their corresponding number of events, $N_{\textrm{C}}$ and $N_{\textrm{bkg}}$, respectively.

The detector resolution function, $G(W',W)$, is observed in data to be non-Gaussian. This is best seen in the distributions of various \Pga peaks, which are instead well-described by a skew-normal distribution with location, width, and skewness parameters $\xi$, $\omega$, and \Pga, respectively:

\begin{equation}
f(x) = \frac{1}{\sqrt{2\pi}\xi}e^{\frac{(x-\xi)^2}{2\omega^2}}\textrm{erfc}\left[\frac{-\alpha(x-\xi)}{\sqrt{2}\omega}\right]
\end{equation}

This description of $G(W',W)$ is accurately predicted by the simulation as a result of variations in light collection efficiency with position. The associated mean and variance of the skew-normal distribution are given by

\begin{align}
\label{eqn:skewmeanvar}
\begin{split}
 \mu &= \xi+\omega\delta\sqrt{\frac{2}{\pi}} ,
\\
 \sigma^2 &= \omega^2\left( 1-\frac{2}{\pi}\delta^2 \right),
 \\
\textrm{with } 
\\
\delta &= \frac{\alpha}{\sqrt{1+\alpha^2}} .
\end{split}
\end{align}

Simulations of uniform energy deposits within the detector volume show that the detector resolution is well modeled by

\begin{equation}
\label{eqn:resModel}
\frac{\sigma}{\mu}=\sqrt{ k_{1}^{2} + \frac{k_{2}^{2}}{\mu} }.
\end{equation}

At higher energies where \Pgg calibration data is available, the first term dominates the resolution behavior. For fitting the low pulse area region, we fix $k_{1}$ to have the value of 0.087 predicted by the Monte Carlo and allow $k_{2}$ to float in the fit.

Extrapolation of the background simulations from Section \ref{ssec:gammaEnhanced} into the low pulse area region indicate that the background shape underlying the \ICof spectrum is well described by a decaying exponential plus a linear term. The integral of this shape then determines the number of background events, $N_{\textrm{bkg}}$, which is fixed during the fit.

The fit is performed between 20 and 80~phe. The following parameters are allowed to float freely:
\begin{itemize}[noitemsep]
\item The number of \ICof events, $N_{\textrm{C}}$
\item The scale factor that converts energy deposited in the LS to the observed number of phe, $Q/E$~(phe/MeV)
\item The constant $k_{2}$ in the detector resolution model, Eq.~(\ref{eqn:resModel})
\item The parameter $a$ in the \Pgb-decay shape factor, $C(W)$.
\end{itemize}

From the fit number of \ICof decays, the concentration \ICof/\ICot is calculated as
\begin{equation}
\label{equ:carbon}
f(^{14}\textrm{C}/^{12}\textrm{C}) = \frac{A \, \tau \, M_{\textrm{LAB}}}{17.1 \, m_{\textrm{LS}} \, N_{A}}
\end{equation}
where $A$ is the fit \ICof activity and $\tau$ is the \ICof mean lifetime (\SI{8266.6}{\year}). $M_{\textrm{LAB}}$ is the molar mass of LAB (233.9~g/mol) and $m_{\textrm{LS}}$ is the mass of LS in the detector during the run. The average number of carbon atoms in LAB is 17.1 and $N_{A}$ is Avogadro's number. Eq.~(\ref{equ:carbon}) returns the ratio \ICof/\ICot as an atom fraction, not a mass fraction\footnotemark.

\footnotetext{Both conventions are used in the literature, often without clear definition of units.}

\begin{figure}[t]
\centering
\includegraphics[width=0.98\textwidth]{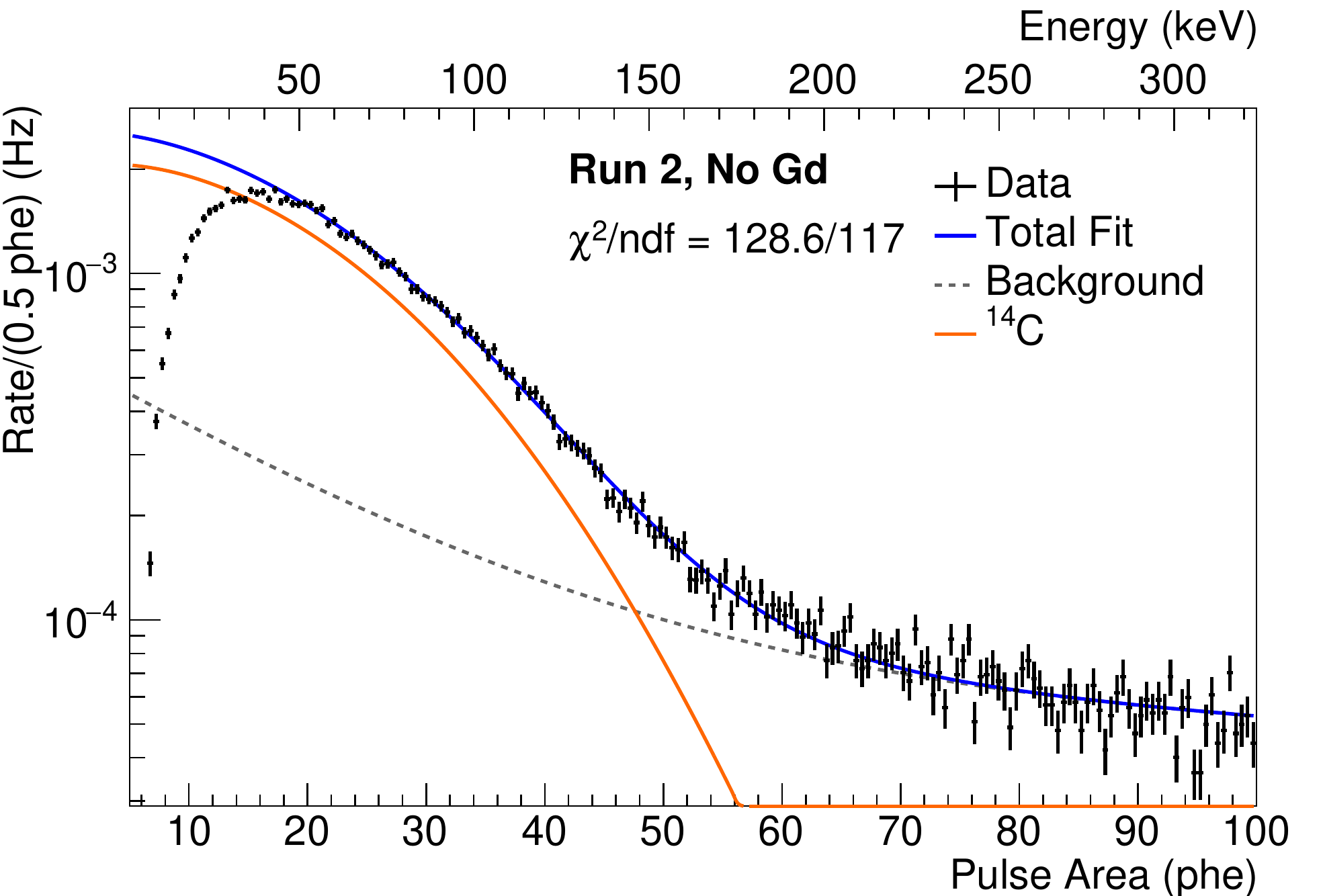}
\caption{\small Run 2 low pulse area data shown with the best fit to the background plus $^{14}\textrm{C}$ \Pgb model. Shown at the top of the plot is the energy axis derived from the best-fit energy scale factor. A full color version of this image is available online.\label{fig:run2Carbon14Fit}}
\end{figure}

A good fit is obtained to the Run 2 data and shown in Fig.~\ref{fig:run2Carbon14Fit}. A summary of the best fit values for the various parameters is given in Tab.~\ref{tab:carbonFitResultsRun2}. The \ICof/\ICot value is given with a statistical error bar from the fit and a systematic error resulting from the uncertainty on the LS mass in the detector. The statistical uncertainty is larger than expected from pure event counting because of correlations between the number of \ICof decays, the energy scale, and energy resolution parameters. Precise knowledge of these parameters in the low-energy region from dedicated calibrations would improve the sensitivity of detectors similar to the Screener.

Our results are consistent with the \ICof/\ICot ratio of $(3.3\pm0.5) \times 10^{-17}$ measured in LAB-based LS in Ref.~\cite{Barabanov:2018hbh}, though our error is approximately a factor of $8\times$ smaller. The level of contamination we measured here satisfies the requirements of the LZ OD. The expected \ICof rate above \SI{100}{\keV} in the OD is \SI{7.0+-0.2}{\Hz}. 

\begin{table}[h]
\centering
\footnotesize
\caption{\small Results of fitting data in the low pulse area region from 20-80~phe with the \ICof + background model in Run 2. \label{tab:carbonFitResultsRun2}}
\begin{tabular*}{0.9\textwidth}{@{\extracolsep{\fill} } l c }
\multicolumn{2}{c}{\textbf{Run 2:} Livetime = \SI{11.83}{\day}; LS Mass = \SI{23.2+-0.1}{\kg}; Fit $\chi^{2}/\textrm{ndf} = 128.6/117$} \\
\hline \hline \textbf{Parameter} & \textbf{Fit Result} \\ \hline
Shape factor, $a$ ($\textrm{MeV}^{-1}$) & $-0.34 \pm 0.12$ \\
Resolution Parameter, $k_{2}$ ($\textrm{phe}^{1/2}$) & $1.57 \pm 0.07$ \\
Energy scale ($\textrm{phe}~\textrm{MeV}^{-1}$) & $309 \pm 4$ \\
\ICof Activity (mBq) & $110.7 \pm 2.4$ \\
Concentration \ICof/\ICot ($\times 10^{-17}$) & $2.83 \pm 0.06 \textrm{(stat.)}\pm 0.01 \textrm{(sys.)}$\\
\hline
\end{tabular*}
\end{table}

\subsection{\label{ssec:bipo}\Pgb-\Pga and \Pga-\Pga Coincidence Rates}

As mentioned in the introduction, the \IUtTeM, \IUtTFL, and \IThtTtL subchains are commonly measured by directly counting \Pgb-\Pga or \Pga-\Pga coincidences. Each of these subchains contains a short-lived polonium isotope which allows a coincidence tag with the previous decay in the series. The relevant polonium half-lives in these three subchains are \SI{164}{\mus} (\IPotof), \SI{1.78}{\ms} (\IPotoF), and \SI{299}{\ns} (\IPotot), respectively. 

Pulse pairs consistent with the expected \Pga and \Pgb energies are counted in two coincidence windows within the larger \SI{81.9}{\mus} event window: \SIrange{5}{76}{\mus} for the \IUtTe and \IUtTF chains and \SIrange{0.4}{4}{\mus} for the \IThtTt chain. The deadtime following each event prevents the use of larger windows for \IPotof and \IPotoF decay. The PSD value of the first pulse in a coincident pair distinguishes the \Pgb-\Pga events from the \IUtTe chain and the \Pga-\Pga events from the \IUtTF chain. The accidental coincidence background is largest when selecting \Pgb-\Pga events in the \IUtTe chain and is calculated to be 0.08 events, making this measurement essentially background-free. Requiring that the second pulse (\Pga pulse) be consistent with the expected quenched alpha energy removes background from PMT afterpulses, as they have much smaller pulse areas~\cite{Baudis:2013xva}.

The efficiency for selecting events with pulse pairs within each timing window is calculated analytically using the half-life of the polonium decay. The efficiency of cuts on pulse area (and therefore deposited energy) is assessed through simulation. The total efficiencies for the \IUtTe, \IUtTF, and \IThtTt chain selections in Run 1 (Run 2) are \SI{21.7}{\percent} (\SI{21.9}{\percent}), \SI{2.7}{\percent} (\SI{2.7}{\percent}), and \SI{32.1}{\percent} (\SI{32.6}{\percent}), respectively. The largest source of inefficiency results from the choice of coincidence windows, a limitation imposed by the maximum event length of the DAQ. The dominant error on the double-pulse efficiency arises from the DAQ resolution in time, which is \SI{10}{\ns}, causing at most an error in efficiency of 0.9\% for the \IThtTt chain selection and 0.01\% for the \IUtTe and \IUtTF chain selections.



The subchain activities measured by coincidence counting in both runs are summarized in Tab.~\ref{tab:bipoResults}. The larger activities in the GdLS are consistent with the expectation that impurities enter through the Gd-loading process. In particular, it was demonstrated in Ref.~\cite{Yeh:2010zz} that the purification process for the Gd compound can efficiently remove thorium, but does not remove actinium. Similar difficulty in removing isotopes of protactinium from Gd compounds was observed in Ref.~\cite{Boiko:2017mmh} due to its chemical similarity with rare-Earth metals. We therefore suppose that \IPatTo and/or \IActtS introduced during Gd-loading serves as the source of the \IUtTFL subchain activity measured here. No events consistent with the \IUtTFL subchain sequence are observed in the unloaded LS data.

\begin{table}[h]
\centering
\footnotesize
\caption{\small Concentrations of various subchains calculated from the measured coincidence rates in each run. \label{tab:bipoResults}}
\begin{tabular}{@{\extracolsep{0.03\textwidth}} c c c} \hline
\textbf{Subchain} & \textbf{Run 1 (mBq/kg)} & \textbf{Run 2 (mBq/kg)}\\ \hline
\IUtTeM  & $0.019 \pm 0.003$ & $0.023 \pm 0.002$ \\
\IUtTFL & $0.18 \pm 0.02$ & $<0.0037$ (90\% CL) \\
\IThtTtL & $0.0071 \pm 0.0019$ & $0.00082 \pm 0.00074$ \\
\hline
\end{tabular}
\end{table}

\subsection{\label{ssec:alphaEnhanced}Fit to \Pga Events}

All of the \Pga decays in the \IUtTe, \IUtTF, and \IThtTt chains contribute to the event rate in the Screener and are distinguished from other sources by using the PSD selection discussed in Sec.~\ref{ssec:psd}. Fits to the pulse area spectra of selected \Pga events are performed to measure the isotope activities throughout these chains.

Probability density functions (PDFs) in pulse area for the primary \Pga-emitting isotopes in the \IUtTe, \IUtTF, and \IThtTt chains are created by simulating 50k decays of each uniformly in the LS volume. The distributions of collected phe are then weighted by the measured phe-dependent acceptance (Fig.~\ref{subfig:alphaeff}) to obtain the expected shapes for each isotope.

Isotopes from the same subchains are summed to reduce the overall number of fit parameters. This results in six isotope populations: \IUtTeE, \IUtTeM, \IPotoz, \IUtTFL, \IThtTt, and \IThtTtL. The \IUtTFL PDF is not used in the Run 2 fit as no signature for this chain was found in Sec.~\ref{ssec:bipo}. The coincidence rates measured in Sec.~\ref{ssec:bipo} constrain their associated subchain PDF normalizations during the fit. The remaining populations float freely.

The large stopping power of \Pga's restricts their origin to the LS itself or the inside acrylic surface of the LS chamber. In the latter case, radon daughter plate-out results in \IPotoz \Pga's depositing their energy in the LS with minimal energy loss in the acrylic due to their \SI{\sim100}{\nm} implantation depth. This is short compared to the typical range of these \Pga's in acrylic which is \SI{\approx30}{\mum}.

An upper limit on the \IPotoz surface activity is known from counting a $\sim$year old sample of acrylic from the manufacturer of the Screener vessel. The sample had no protective coating during its exposure time. An XIA UltraLo-1800 \Pga counter~\cite{XIA:2018} measured a \IPotoz surface activity of \SI{3.0+-0.7}{\mBq\per\m^2} on this sample. The inside surface area of the Screener LS chamber is \SI{0.52}{\m^2}, while only half the \IPotoz \Pga's interact in the LS. This gives the conservative upper limit of \SI{0.8+-0.2}{\mHz} in the Screener. Given that the Screener is constructed of virgin acrylic and was kept in a nitrogen atmosphere whenever possible, the true surface activity is likely much smaller and so is ignored here.

\subsection{\label{ssec:gammaEnhanced}Fit to \Pgg/\Pgb Events}

Complementary fits are performed to spectra of \Pgg/\Pgb events selected using the cut described in Sec.~\ref{ssec:psd}. Subchain equilibrium suggests that activity measured through \Pga-decays should have a concomitant activity through \Pgb-decays. The populations included in these fits are summarized here.

Event distributions resulting from sources of radioactivity external to the LS are obtained by simulation of decays in the detector construction materials. The expected number of counts from external \IUtTe, \IThtTt, and \IKfz are then obtained by normalizing to the detector material radioassay results from HPGe counting. The contribution from \IRnttt decays in the surrounding water is also included. The rate from \ICosz in the detector materials was found to be negligible.

In the \IUtTe, \IUtTF, and \IThtTt chains, there are six, four, and five primary isotopes that decay by \Pgb emission, respectively. Of these, $^{234}\textrm{Th}$, $^{210}\textrm{Pb}$, $^{227}\textrm{Ac}$, $^{231}\textrm{Th}$, and $^{228}\textrm{Ra}$ are excluded as their \Pgb-decay endpoints fall below the range of the fit. As was done in the \Pga fits, isotopes belonging to the \IUtTeM, \IUtTFL, and \IThtTtL subchains are summed and constrained by the coincidence rates measured in Sec.~\ref{ssec:bipo}. The isotopes that remain are allowed to float free, namely: $^{234}\textrm{Pa}$, \IBitoz, $^{228}\textrm{Ac}$.

The cosmic ray flux of protons and neutrons on the Earth's surface leads to the production of the isotope $^{7}\textrm{Be}$ through reactions on \ICot in the LS. The resulting electron capture decay of $^{7}\textrm{Be}$ can enter as a source of background through its emission of a \SI{478}{\keV} \Pgg-ray \SI{10.4}{\percent} of the time. The \SI{53}{\day} half-life of $^{7}\textrm{Be}$ means that the surface production rate and decay rate were very likely in equilibrium for both the loaded and unloaded LS used here. The expected rate of \SI{478}{\keV} \Pgg-rays from $^{7}\textrm{Be}$ in the pseudocumene-based LS used by Borexino was estimated in~\cite{Vogelaar:1996a}. By scaling those results and accounting for time spent underground (about \SI{48}{\day} and \SI{29}{\day} in Run 1 and 2, respectively), we might expect \SI{\approx1}{\mBq} of \Pgg-emitting $^{7}\textrm{Be}$ decays in the low-background data from each run. An independent estimate performed using the \textsc{ACTIVIA} software package~\cite{back:2007kk} predicts a rate about a factor of $3\times$ higher (demonstrating the large uncertainty associated with the activation rates). A PDF for $^{7}\textrm{Be}$ decay is included in the fits for both runs.

Another possible contaminant is \IKreF. Present at the level of \SI{\sim1}{\Bq/\m^3} in atmospheric air, \IKreF decays primarily by emission of a \SI{252}{\keV} \Pgb with a half-life of \SI{10.7}{\year}. In \SI{0.434}{\percent} of these decays, the daughter nucleus is left in a meta-stable state with half-life \SI{1.01}{\mus}. The deexcitation of this state results in the emission of a \SI{514}{\keV} \Pgg that can be tagged in coincidence with the \Pgb. A set of selection cuts similar to those in Sec.~\ref{ssec:bipo} were devised to search for these coincidences in each run with total cut efficiencies of \SI{13.6}{\percent} and \SI{11.7}{\percent} with errors of at most 0.5\%. Two events pass these cuts in Run 1, while a single event passes in Run 2. The expected backgrounds from accidental coincidences are 0.035 and 0.003 events. The resulting concentrations fall within the intervals $[0.07,0.81]$~mBq/kg and $[0.01,0.36]$~mBq/kg at \SI{90}{\percent} CL, respectively. These are consistent with estimates of the air exposure during filling, which make use of the initial rates of \IRnttt in each run. \IKreF populations in each fit are constrained with these normalization intervals. In the much larger OD, preventing \IKreF contamination during LS filling will be crucial.


In summary, the following populations are included with constraint terms in the fit: external U, Th, K, water \IRnttt decays, \IUtTeM, \IUtTFL (Run 1 only), \IThtTtL, \ILuoSs (Run 1 only), and \IKreF. The following populations are included and allowed to float freely: $^{234}\textrm{Pa}$, $^{210}\textrm{Bi}$, $^{228}\textrm{Ac}$, $^{7}\textrm{Be}$, and $^{40}\textrm{K}$.

\subsection{Fit Results\label{ssec:results}}

Our strategy is to regard the \Pga fits as the most robust technique for measuring the scintillator impurities. This comes about as a result of two observations: first, the shapes resulting from \Pga events are more distinct than those from \Pgg/\Pgb events, which tend to be relatively featureless and can overlap significantly. Second, the \Pgg/\Pgb fits suffer from having the external activity of the detector components as a background and contain a larger number of fit parameters.

The results from all fits are summarized in Tab.~\ref{tab:results}, where isotopes are grouped according to their decay chain. Where subchain results are given, the specified activity represents the activity of each isotope within the subchain. In general, we find consistency between the \Pga fits and the \Pgg/\Pgb fits within the various subchains where equilibrium is expected. Activity is measured in the early parts of both the \IUtTe and \IThtTt chains different than that measured by \Pgb-\Pga coincidences, demonstrating that secular equilibrium is broken. The elevated activity of the GdLS with respect to the unloaded LS sample is obvious. The inclusive Screener rates above \SI{200}{\keV} reconstructed from Tab.~\ref{tab:results} are \SI{89.5+-7.3}{\mHz} in Run 1 and \SI{25.1+-3.0}{\mHz} in Run 2, which are generally consistent (within $1.7\sigma$) with those measured in data.

The isotope concentrations measured in the Screener generally agree with those obtained by the HPGe assay in Tab.~\ref{tab:gdScreening}. The largest discrepancy between those results and the Screener results is in the \IThtTtE subchain concentration. In the HPGe analysis this subchain is measured using \Pgg lines following the decay of \IActte. The sensitivity of the Screener to this \Pgb-emitting isotope is poor, however the Screener is able to detect the \Pga-decays from \IThtTt at the head of the subchain directly. As the Screener and HPGe detectors measure different isotopes in the subchain, the discrepancy between the two may be explained by a breakage of subchain equilibrium at the long-lived \IRatte isotope which has a half-life of \SI{5.8}{\year}.



\begin{figure}[h]
\centering
\begin{subfigure}{0.95\textwidth}
\includegraphics[width=\textwidth]{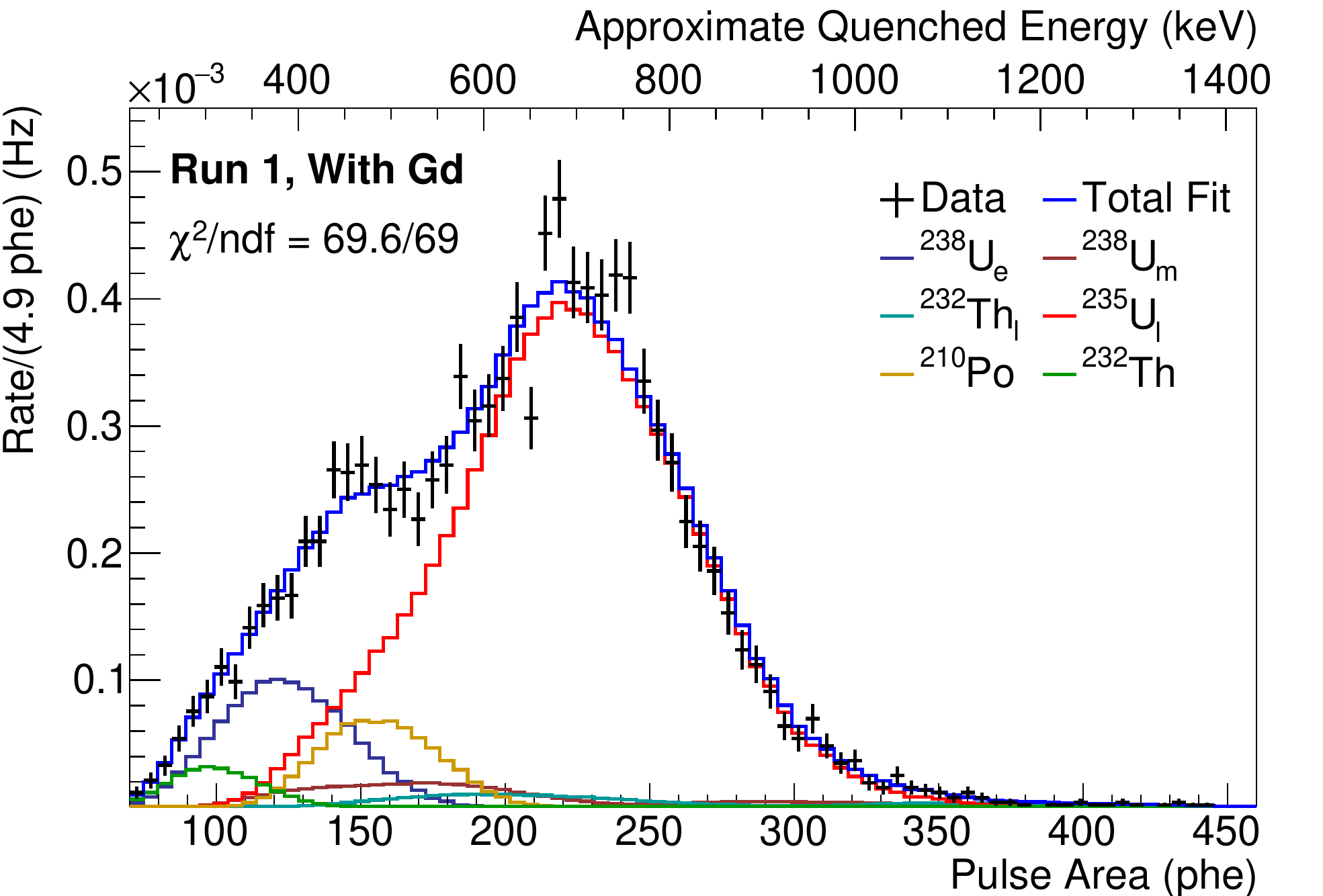}
\subcaption{\label{subfig:run1alpha}}
\end{subfigure}
\begin{subfigure}{0.95\textwidth}
\includegraphics[width=\textwidth]{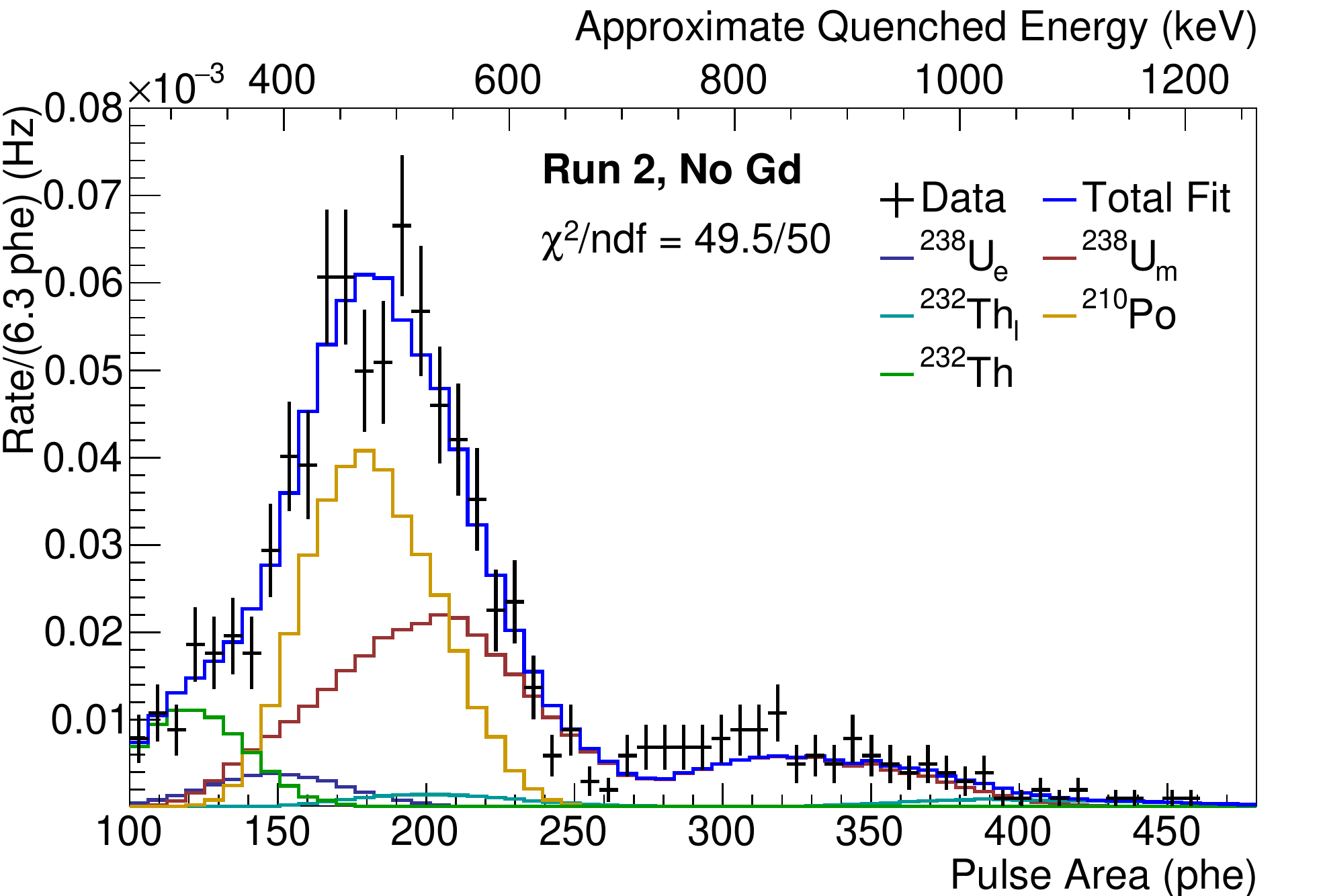}
\subcaption{\label{subfig:run2alpha}}
\end{subfigure}
\caption{\small Best-fit to selection of \Pga events in (a) the GdLS sample and (b) the unloaded LS sample. A full color version of this image is available online. \label{fig:alphaFits}}
\end{figure}

The \Pga fits are shown in Fig.~\ref{fig:alphaFits}. In the unloaded LS of Run 2, the prominent feature is from out-of-equilibrium \IPotoz decay. In the GdLS, the \Pga rate is mostly comprised of decays from the \IUtTFL subchain, resulting in a smaller error on its concentration than that obtained from \Pga-\Pga coincidences.

\begin{figure}[h]
\centering
\begin{subfigure}{0.95\textwidth}
\includegraphics[width=\textwidth]{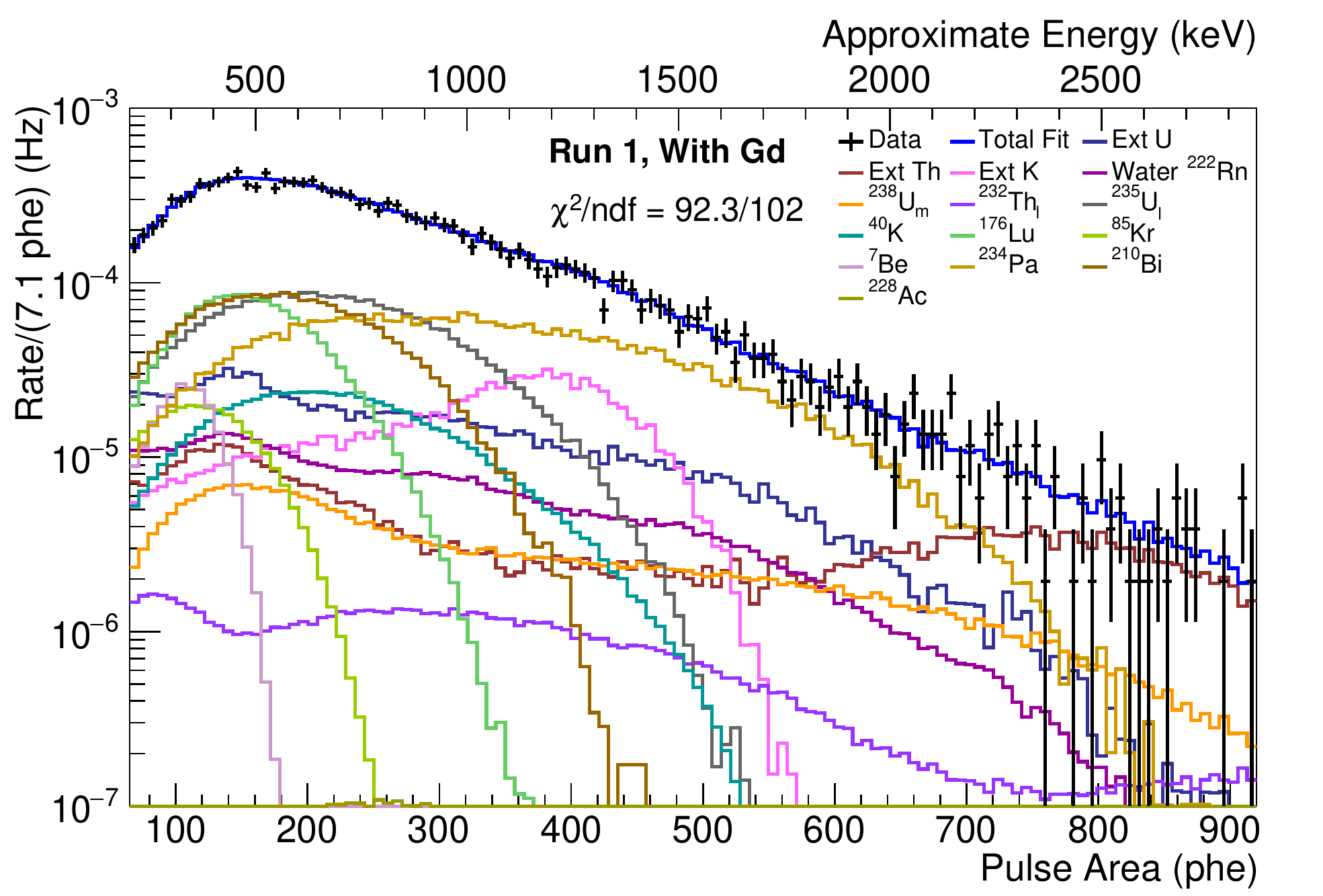}
\subcaption{\label{subfig:run1gamma}}
\end{subfigure}
\begin{subfigure}{0.95\textwidth}
\includegraphics[width=\textwidth]{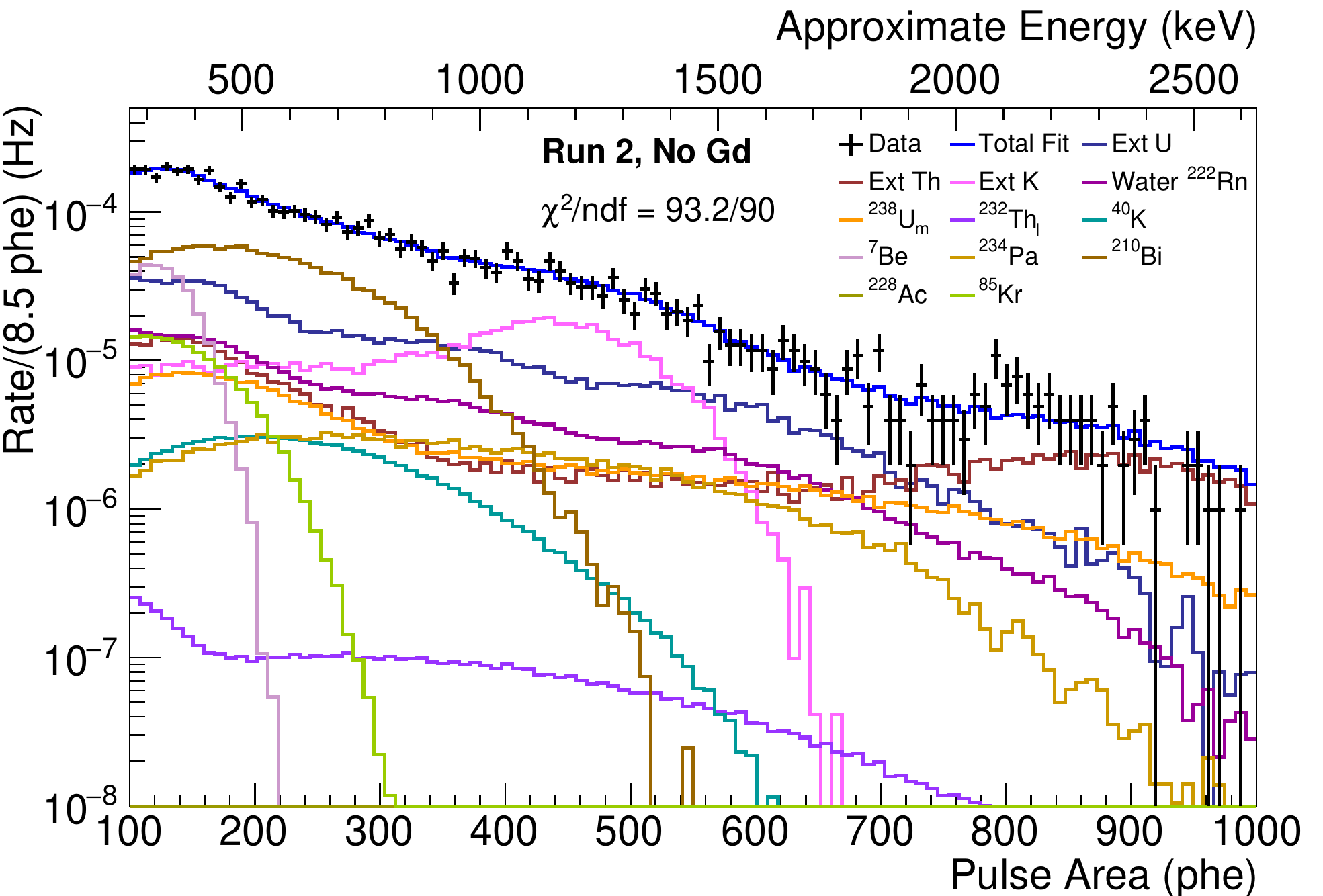}
\subcaption{\label{subfig:run2gamma}}
\end{subfigure}
\caption{\small Best-fit to selection of \Pgg/\Pgb events in (a) the GdLS sample and (b) the unloaded LS sample. A full color version of this image is available online. \label{fig:gammaFits}}
\end{figure}

The \Pgg/\Pgb fits are shown in Fig.~\ref{fig:gammaFits} and provide a useful cross-check for the \Pga results. In the Run 2 \Pgg/\Pgb fit the best-fit values for the external uranium, external potassium, and water \IRnttt activities are in general agreement with their predicted values. The amount of external thorium found by the fit is roughly a factor of two lower than predicted, but well constrained by the highest pulse area data. As the external conditions in each run were the same, we use the best-fit value found in Run 2 to rescale the external thorium constraint for the GdLS fit in Run 1.

\begin{table}[h]
\centering
\footnotesize
\caption{\small Measured radioimpurity concentrations in both loaded and unloaded LS runs grouped by isotope type. Reported errors are statistical only and, where none are given, the one-sided upper limit at 90\% CL is reported. A 0.4\% systematic error resulting from the uncertainty on the LS mass in the detector also applies. \label{tab:results}}
\noindent\adjustbox{max width=\textwidth}{
\begin{tabular}{ l l c c c }
\hline
\multicolumn{2}{c}{\multirow{2}{*}{\textbf{Isotope}}} & \textbf{Gd-Loaded LS Activity} &  \textbf{Unloaded LS Activity} & \multirow{2}{*}{\textbf{Method}} \\ 
 & & \textbf{(mBq/kg)} &  \textbf{(mBq/kg)} & \\
\hline
\multirow{6}{*}{\textbf{$^{238}\textrm{U}$ Chain}} & $^{238}\textrm{U}$,$^{234}\textrm{U}$  & $0.23 \pm 0.02$ & - & \Pga Fit \\
 & $^{238}\textrm{U}$,$^{234}\textrm{U}$,$^{230}\textrm{Th}$  & - & $0.0055\pm0.0052$ & \Pga Fit \\
& $^{234}\textrm{Pa}$ & $0.33 \pm 0.04$  & $<0.065$        & $\gamma/\beta$ Fit \\
& \IUtTeM Subchain  & $0.019 \pm 0.003$ & $0.023 \pm 0.002$ & BiPo \\
& $^{210}\textrm{Bi}$ & $0.40 \pm 0.26$ & $0.30 \pm 0.10$ & $\gamma/\beta$ Fit \\
& $^{210}\textrm{Po}$ & $0.16 \pm 0.02$ & $0.099 \pm 0.009$ & \Pga Fit \\
\hline
\multirow{1}{*}{\textbf{$^{235}\textrm{U}$ Chain}} & \IUtTFL Subchain & $0.185 \pm 0.006$ & - & \Pga Fit \\
\hline
\multirow{3}{*}{\textbf{$^{232}\textrm{Th}$ Chain}} & $^{232}\textrm{Th}$ & $0.16 \pm 0.04$ & $0.059 \pm 0.013$ & \Pga Fit \\
 & $^{228}\textrm{Ac}$ & $< 0.20$         & $< 0.018$          & $\gamma/\beta$ Fit \\
 & \IThtTtL Subchain & $0.0071 \pm 0.0019$ & $0.00082 \pm 0.00074$ & BiPo \\
\hline
\multirow{4}{*}{\textbf{Other}} & $^{40}\textrm{K}$   & $<0.34$  & $<0.14$  & $\gamma/\beta$ Fit \\
 & $^{7}\textrm{Be}$   & $<2.69$  & $1.67 \pm 0.51$ & $\gamma/\beta$ Fit \\
  & $^{85}\textrm{Kr}$   & $<0.31$  & $0.069 \pm 0.067$ & $\gamma/\beta$ Fit \\
 & $^{176}\textrm{Lu}$ & $0.25 \pm 0.07$  & - & $\gamma/\beta$ Fit \\
\hline
\end{tabular}}
\end{table}

Our hypotheses regarding the locations of decay chain equilibrium breakage are motivated by the purification methods used to prepare the LS and GdLS discussed in Sec.~\ref{sec:LSprops}. Both the water extraction and distillation techniques have demonstrated efficient removal of heavy elements, in particular, lead and radium~\cite{Ford:2011zza,Keefer:2013eaa}. In comparison, the Gd compound's purification effectively removes thorium, but is not effective at removing actinium as previously discussed in Sec.~\ref{ssec:bipo}. Our hypotheses are therefore the following: in the unloaded LS sample of Run 2 we expect that equilibrium is broken at the long-lived radium isotopes \IRatts (\Pga emitter, \IUtTe chain) and \IRatte (soft \Pgb emitter, \IThtTt chain). In the GdLS, we expect that efficient thorium removal leads to breaks at the long-lived thorium isotopes \IThtTz (\Pga emitter, \IUtTe chain) and \IThtte (\Pga emitter, \IThtTt chain).

The most sensitive test for finding the location of equilibrium breakage is in the \Pga fits for each run. The \Pga emitters \IUtTe through \IRatts in the \IUtTe chain are successively removed from the BiPo-constrained \IUtTeM population and allowed to float freely, both independently from one another and as a summed \IUtTeE population. The fit $\chi^{2}$/ndf from each iteration is recorded in Tab.~\ref{tab:brokenChain}. Improvement in the Run 1 fit is observed with the removal of isotopes up to \IThtTz, after which the improvement is very small. A similar improvement is observed in Run 2, however, the case where no early \IUtTe chain isotopes are excluded from the \IUtTeM population yields an acceptable $\chi^{2}$ at \SI{90}{\percent} CL. This is reflected in Tab.~\ref{tab:results} where the concentration of \IUtTe+\IUtTf is consistent with the measured BiPo coincidence rate.

\begin{table}[h]
\centering
\footnotesize
\caption{\small $\chi^{2}$ values obtained when various $^{238}\textrm{U}$ early chain isotopes are allowed to float free, independent from the BiPo-constrained, \IUtTeM PDF ending with $^{214}\textrm{Po}$. Also given are the $\chi^{2}$ values when those early chain isotopes are summed into a single, early chain PDF.\label{tab:brokenChain}}
\begin{tabular}{@{\extracolsep{0.03\textwidth}} l c c c c} \hline
\multirow{2}{*}{\textbf{Early $^{238}\textrm{U}$ Chain Isotopes}} & \multicolumn{2}{c}{\textbf{Run 1 $\chi^{2}/\textrm{ndf}$}} & \multicolumn{2}{c}{\textbf{Run 2 $\chi^{2}/\textrm{ndf}$}} \\ 
 & \textbf{Free} & \textbf{Summed} & \textbf{Free} & \textbf{Summed} \\ \hline
None & \multicolumn{2}{c}{147.4/70} & \multicolumn{2}{c}{63.2/51} \\
$^{238}\textrm{U}$ & 95.9/69 & 95.9/69 & 58.4/50 & 58.4/50 \\
$^{238}\textrm{U}$, $^{234}\textrm{U}$ & 68.1/68 & 69.6/69 & 50.4/49 & 50.4/50 \\
$^{238}\textrm{U}$, $^{234}\textrm{U}$, $^{230}\textrm{Th}$ & 68.0/67 & 69.1/69 & 49.5/48 & 49.5/50 \\
$^{238}\textrm{U}$, $^{234}\textrm{U}$, $^{230}\textrm{Th}$, $^{226}\textrm{Ra}$ & 68.0/66 & 68.9/69 & 49.1/47 & 49.4/50 \\
\hline
\end{tabular}
\end{table}

\FloatBarrier

\subsection{Inclusive Fits and Results}

The inclusive Screener count rates above \SI{200}{\keV} implied by the radioimpurity concentrations of the previous section are within $1.7\sigma$ of that observed in data. An additional inclusive fit was performed in each run to observe how the above results change under the constraint of equal rates, with no selection on pulse shape imposed. During the fit, all isotope populations begin with the concentrations measured in Tab.~\ref{tab:results} and are constrained by those errors as well. For each of the \IUtTeM, \IUtTFL, and \IThtTtL subchains, the \Pga and \Pgb emitting isotopes are summed into single populations. Furthermore, to reduce the number of fit parameters, the \IUtTeM and \IThtTtL populations are fixed, as their rates are subdominant and have already been measured by counting pulse pairs in Sec.~\ref{ssec:bipo}. The \IUtTFL population is allowed to float as the combined rate from its \Pga activity forms a very prominent feature in the pulse area spectrum of Run 1.

\begin{figure}[h]
\centering
\begin{subfigure}{0.95\textwidth}
\includegraphics[width=\textwidth]{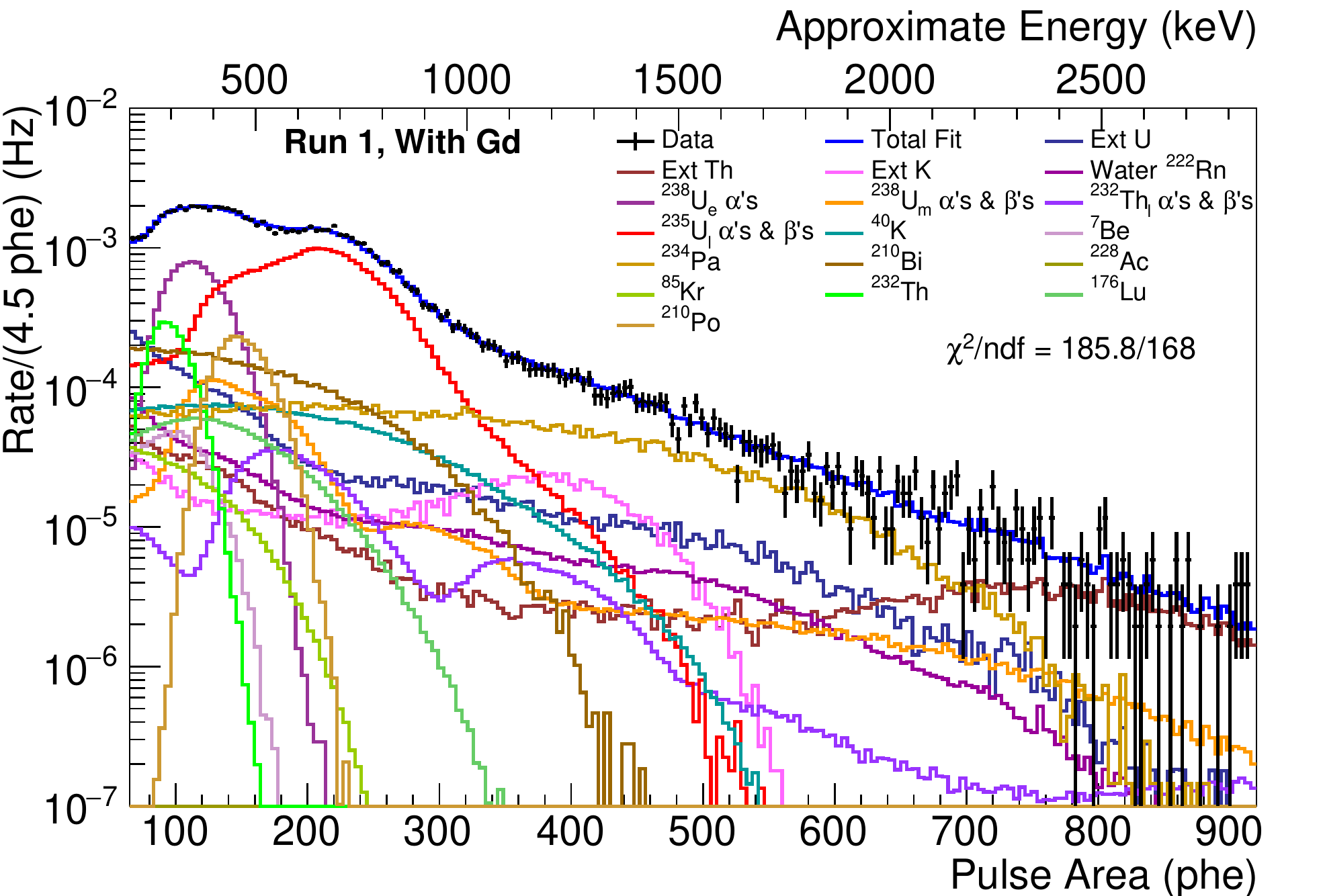}
\end{subfigure}
\begin{subfigure}{0.95\textwidth}
\includegraphics[width=\textwidth]{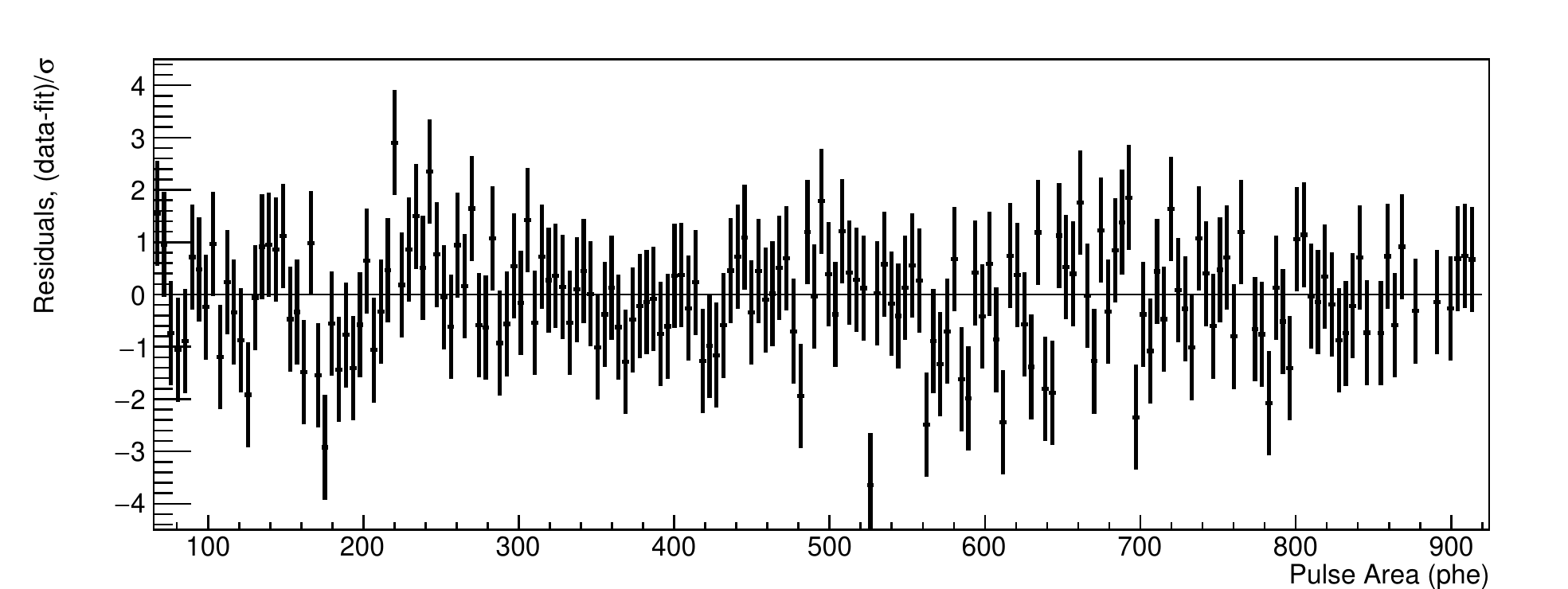}
\end{subfigure}
\caption{\small Inclusive fit to Run 1 data and fit residuals. A full color version of this image is available online.\label{fig:inclusiveFitRun1}}
\end{figure}

\begin{figure}[h]
\centering
\begin{subfigure}{0.95\textwidth}
\includegraphics[width=\textwidth]{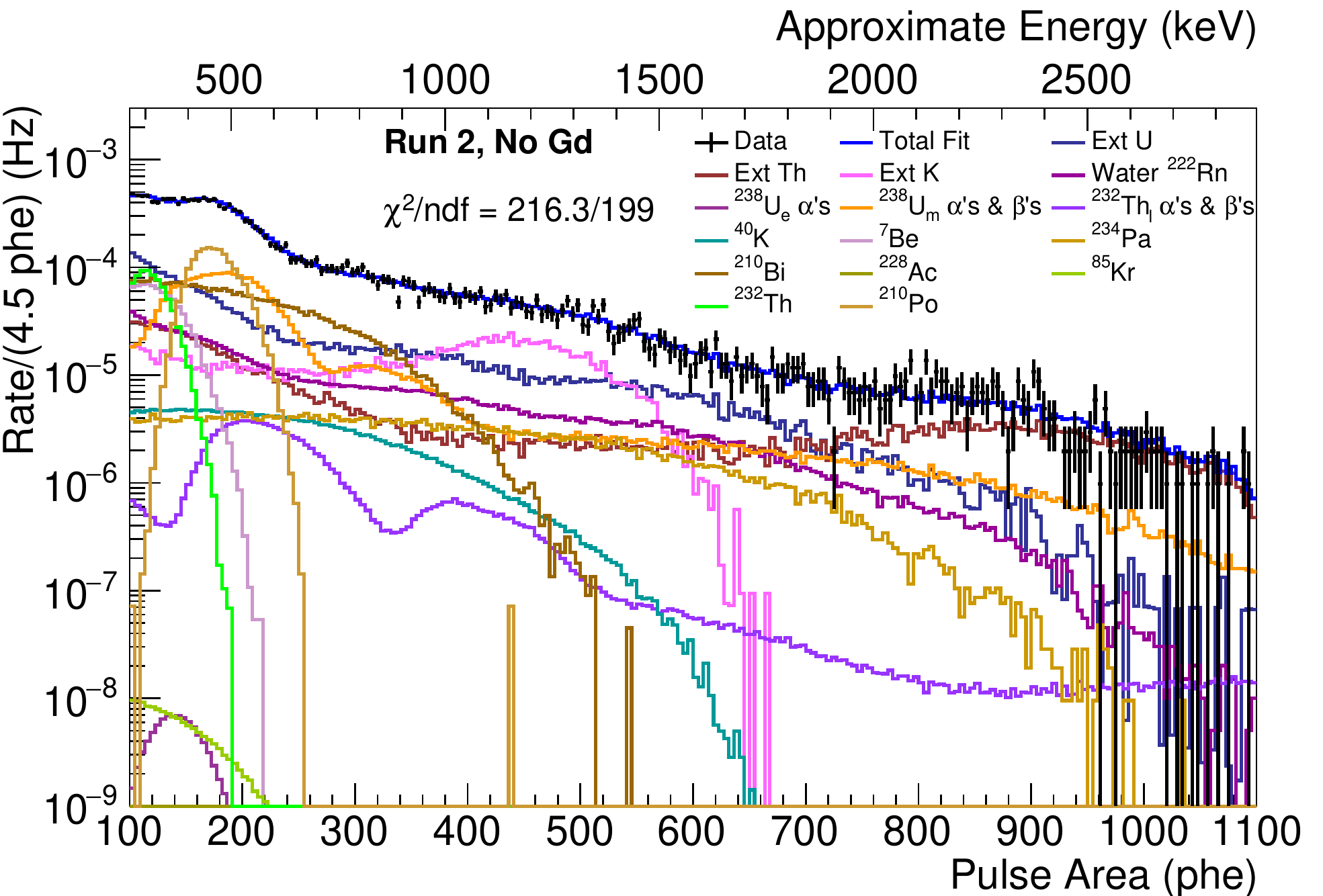}
\end{subfigure}
\begin{subfigure}{0.95\textwidth}
\includegraphics[width=\textwidth]{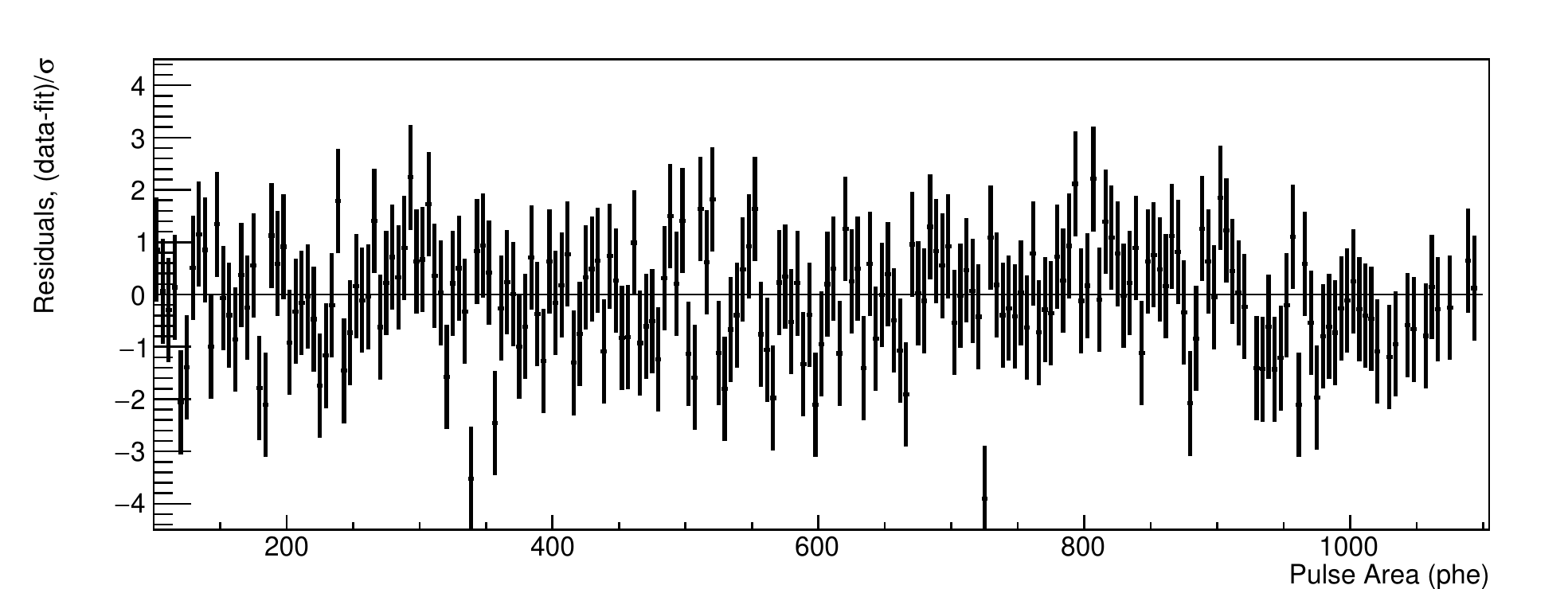}
\end{subfigure}
\caption{\small Inclusive fit to Run 2 data and fit residuals. A full color version of this image is available online.\label{fig:inclusiveFitRun2}}
\end{figure}

The inclusive fit results are given in Tab.~\ref{tab:inclusive_results} where the results of the previous section are also repeated for comparison. The fits and their residuals are shown in Fig.~\ref{fig:inclusiveFitRun1} and Fig.~\ref{fig:inclusiveFitRun2}. Generally, we find a uniform reduction in concentration across all the isotope populations. We note, however, the $2.6\sigma$ discrepancy between results for $^{176}\textrm{Lu}$ in Run 1. Two attempts were made to rectify these results. First, the $^{176}\textrm{Lu}$ rate was fixed, which resulted in a change of $\Delta \chi^2=9.0$. Second, the constraint on $^{210}\textrm{Bi}$ was removed as its shape is similar to that of $^{176}\textrm{Lu}$. In this iteration the best fit $^{176}\textrm{Lu}$ and $^{210}\textrm{Bi}$ rates were the same as that reported here.

\begin{table}[h]
\centering
\footnotesize
\caption{\small Radioimpurity concentrations measured in the \Pga \& \Pgb/\Pgg fits and using the inclusive fit in both loaded and unloaded LS runs grouped by isotope type. Reported errors are statistical only and, where none are given, the one-sided upper limit at 90\% CL is reported. A 0.4\% systematic error resulting from the uncertainty on the LS mass in the detector also applies. \label{tab:inclusive_results}}
\noindent\adjustbox{max width=\textwidth}{
\begin{tabular}{ l l c c c c }
\hline
\multicolumn{2}{c}{\multirow{2}{*}{\textbf{Isotope}}} & \multicolumn{2}{c}{\textbf{Gd-Loaded LS Activity (mBq/kg)}} &  \multicolumn{2}{c}{\textbf{Unloaded LS Activity (mBq/kg)}} \\
 & & \textbf{$\alpha$ \& $\gamma$/$\beta$ Fit} & \textbf{Inclusive Fit} & \textbf{$\alpha$ \& $\gamma$/$\beta$ Fit} & \textbf{Inclusive Fit} \\ \hline

\multirow{6}{*}{\textbf{$^{238}\textrm{U}$ Chain}} & $^{238}\textrm{U}$,$^{234}\textrm{U}$  & $0.23 \pm 0.02$ & $0.20 \pm 0.01$ & - & - \\
 & $^{238}\textrm{U}$,$^{234}\textrm{U}$,$^{230}\textrm{Th}$  & - & - & $0.0055\pm0.0052$ & $<1.3$ \\
& $^{234}\textrm{Pa}$ & $0.33 \pm 0.04$ & $0.31 \pm 0.02$ & $<0.065$ & $0.021\pm0.007$ \\
& \IUtTeM Subchain  & \multicolumn{2}{c}{$0.019 \pm 0.003$} & \multicolumn{2}{c}{$0.023 \pm 0.002$}  \\
& $^{210}\textrm{Bi}$ & $0.40 \pm 0.26$ & $0.39 \pm 0.12$ & $0.30 \pm 0.10$ & $0.19 \pm 0.01$ \\
& $^{210}\textrm{Po}$ & $0.16 \pm 0.02$ & $0.11 \pm 0.02$ & $0.099 \pm 0.009$ & $0.08 \pm 0.01$ \\
\hline
\multirow{1}{*}{\textbf{$^{235}\textrm{U}$ Chain}} & \IUtTFL Subchain & $0.185 \pm 0.006$ & $0.19 \pm 0.02$ & - & - \\
\hline
\multirow{3}{*}{\textbf{$^{232}\textrm{Th}$ Chain}} & $^{232}\textrm{Th}$ & $0.16 \pm 0.04$ & $0.10 \pm 0.02$ & $0.059 \pm 0.013$ & $0.037 \pm 0.008$ \\
 & $^{228}\textrm{Ac}$ & $< 0.20$ & $< 0.08$ & $< 0.018$ & $< 0.010$ \\
 & \IThtTtL Subchain & \multicolumn{2}{c}{$0.0071 \pm 0.0019$} & \multicolumn{2}{c}{$0.00082 \pm 0.00074$} \\
\hline
\multirow{4}{*}{\textbf{Other}} & $^{40}\textrm{K}$   & $<0.34$ & $0.20 \pm 0.08$ & $<0.14$ & $0.015\pm0.008$ \\
 & $^{7}\textrm{Be}$   & $<2.69$  & $<2.93$ & $1.67 \pm 0.51$ & $1.16 \pm 0.59$ \\
  & $^{85}\textrm{Kr}$   & $<0.31$  & $<0.29$ & $0.069 \pm 0.067$ & $<0.02$ \\
 & $^{176}\textrm{Lu}$ & $0.25 \pm 0.07$ & $0.08 \pm 0.06$ & - & - \\
\hline
\end{tabular}}
\end{table}

\FloatBarrier

\subsection{Rates in the OD}

In Tab.~\ref{tab:ODrate} we use the Screener results from Tab.~\ref{tab:results} to predict the contributions to the singles rate in the \SI{17.3}{\tonnesl} of LS in the LZ OD. Rates are reported separately for \Pga and \Pgb/\Pgg decays and also given assuming no energy threshold and a \SI{100}{\keV} threshold. The contribution from \IBeS is not included as its half-life is small compared to the planned length of the experiment~\cite{Akerib:2018lyp}. Similar results for the total \Pga and \Pgb/\Pgg rates in the OD derived from the inclusive fits given Tab.~\ref{tab:inclusive_results} are consistent within the errors given here.

\begin{table}[h]
\centering
\footnotesize
\caption{\small Predictions for the singles rate in \SI{17.3}{\tonnesl} of LS in the OD calculated from the values in Tab.~\ref{tab:results}. Rates are given in the case of no energy threshold and a \SI{100}{\keV} threshold for both the unloaded and Gd-loaded LS. \label{tab:ODrate}}
\noindent\adjustbox{max width=\textwidth}{
\begin{tabular}{@{\extracolsep{0.03\textwidth}} l l c c c c }
\hline
\multirow{2}{*}{\textbf{Source}} & \multirow{2}{*}{\textbf{Decay Type}} & \multicolumn{2}{c}{\textbf{Gd-Loaded LS Rate~(Hz)}} &  \multicolumn{2}{c}{\textbf{Unloaded LS Rate~(Hz)}} \\
 & & \textbf{No Threshold} &  \textbf{\SI{>100}{\keV}} & \textbf{No Threshold} & \textbf{\SI{>100}{\keV}} \\ \hline
\multirow{3}{*}{\textbf{$^{238}\textrm{U}$ Chain}} & \Pga & $12.4\pm0.6$ & $12.4\pm0.6$ & $3.6\pm0.2$ & $3.6\pm0.2$ \\
 & \Pgb/\Pgg & $22.5\pm6.4$ & $11.9\pm4.0$ & $11.4\pm2.5$ & $5.4\pm1.5$ \\
 & All & $34.8\pm6.4$ & $24.3\pm4.0$ & $15.0\pm2.5$ & $9.0\pm1.5$ \\
 \hline
\multirow{3}{*}{\textbf{$^{235}\textrm{U}$ Chain}} & \Pga & $19.2\pm0.3$ & $19.2\pm0.3$ & 0 & 0 \\
 & \Pgb/\Pgg & $9.6\pm0.2$ & $5.8\pm0.1$ & 0 & 0 \\
 & All & $28.8\pm0.3$ & $25.0\pm0.3$ & 0 & 0 \\
 \hline
\multirow{3}{*}{\textbf{$^{232}\textrm{Th}$ Chain}} & \Pga & $3.4\pm0.7$ & $3.4\pm0.7$ & $1.1\pm0.2$ & $1.1\pm0.2$ \\
 & \Pgb/\Pgg & $0.2\pm3.9$ & $0.2\pm2.7$ & $0.0\pm0.3$ & $0.0\pm0.2$ \\
 & All & $3.6\pm4.0$ & $3.6\pm2.8$ & $1.1\pm0.4$ & $1.1\pm0.3$ \\
 \hline
\textbf{$^{40}\textrm{K}$}   & \Pgb/\Pgg & $1.7\pm3.1$ & $1.6\pm2.9$ & $0.3\pm1.7$ & $0.3\pm1.6$ \\
\textbf{$^{85}\textrm{Kr}$}  & \Pgb/\Pgg & $1.9\pm2.8$ & $1.5\pm2.2$ & $1.2\pm1.2$ & $0.9\pm0.9$ \\
\textbf{$^{176}\textrm{Lu}$} & \Pgb/\Pgg & $4.3\pm1.2$ & $4.3\pm1.2$ & 0 & 0 \\
\textbf{$^{14}\textrm{C}$}   & \Pgb & $82.5\pm1.7$ & $7.0\pm0.2$ & $82.5\pm1.7$ & $7.0\pm0.2$ \\
\textbf{$^{152}\textrm{Gd}$} & \Pga & $27.9\pm1.4$ & $17.0\pm0.9$ & 0 & 0 \\
\textbf{$^{147}\textrm{Sm}$} & \Pga & $18.2\pm1.8$ & $13.5\pm1.4$ & 0 & 0 \\
\hline
\textbf{Total} & \Pga & $81.1\pm2.5$ & $65.5\pm1.9$ & $4.7\pm0.3$ & $4.7\pm0.3$  \\
\textbf{Total} & \Pgb/\Pgg & $122.8\pm8.8$ & $32.5\pm6.1$ & $95.4\pm3.7$ & $13.7\pm2.4$ \\
\textbf{Total} & All & $203.9\pm9.2$ & $97.9\pm6.4$ & $100.1\pm3.7$ & $18.3\pm2.4$ \\
\hline
\end{tabular}}
\end{table}

In the OD, \Pn's which moderate and subsequently capture in the GdLS can be tagged by the unique combination of a prompt, quenched proton recoil signal followed by a delayed electron recoil signal from the cascade of \Pn-capture \Pgg's. In the cases were the proton recoil signal is below threshold or non-existent, internal \Pgb/\Pgg decays can mimic the delayed \Pn-capture signature. Internal \Pga-decays, however, can be identified as being \Pga-like by their pulse shape and so contribute less to the overall false veto probability. Use of a \SI{200}{\keV} threshold in the OD additionally reduces the rate from radioimpurities to \SI{56.8+-5.5}{\Hz} primarily by excluding pulses from \ICof, \IGdoFt, and \ISmofS.

The overall cleanliness of the OD will be higher than that achieved in runs with the Screener. In particular, we expect the rate from \IPbtoz daughters to be reduced through minimized exposure of OD components with radon-laden air. This includes not only the acrylic tanks themselves but also the inside surfaces of the LS filling system. Further improved filling techniques will also minimize the concentration of \IKreF and \IRnttt in the final GdLS. With these considerations the GdLS mixture measured here should be suitable for use in the LZ OD.

\section{Conclusion}
Efficient tagging of background \Pn's in LZ will be accomplished with the use of an OD containing an LS mixture loaded with Gd at \SI{0.1}{\percent} by mass. The LZ requirement for a false veto rate \SI{<5}{\percent} imposes a limit on the concentration of radioimpurities that can contribute to the OD singles rate above a \SI{100}{\keV} threshold. The average impurity level should be \SI{\lesssim0.07}{\mBqkg}.

A small acrylic detector was constructed and operated in the extremely low-background environment of the former LUX water tank to assess the radiopurity of the GdLS. The use of radiopure materials and the very low-background R11410-20 PMTs allows for \SI{\approx e-4}{\mBqkg} sensitivity to radioimpurities in the LS. The detector was used to count samples of both Gd-loaded and unloaded LS. A higher activity from internal radioimpurities is detected in the loaded LS, in agreement with the general experience of experiments seeking to detect neutrinos and antineutrinos.

PSD was used to separately select events from \Pga's and those from \Pgg/\Pgb's. The fits to the resulting pulse area spectra provide a measurement of the concentrations of various radioisotopes in the LS samples. In particular, the detector is sensitive to the rate of \Pga-decays throughout the \IUtTe, \IUtTF, and \IThtTt decay chains. Our results demonstrate that secular equilibrium is broken in all three chains. In the GdLS, the dominant activity is found to be from the $^{231}\textrm{Pa}$ or $^{227}\textrm{Ac}$ subchain of the \IUtTF series. The strong \Pgg-line from \IUtTF at the head of this decay chain was not detected in a HPGe screening of the purified Gd additives used in the loading process. The HPGe assay results generally agree within errors with those obtained in the Screener. The discrepancy in the measured \IThtTtE concentrations is possibly explained by removal of \IRatte during the Gd additive purification.

The \ICof/\ICot ratio of the unloaded LS was measured to be $(2.83\pm0.06\textrm{(stat.)}\pm0.01\textrm{(sys.)})\times10^{-17}$, the sensitivity of our detector being comparable to that of detectors two orders of magnitude more massive.

The expected rate in the LZ OD from GdLS radioimpurities derived using the Screener measurements is $97.9\pm6.4$~Hz above a  \SI{100}{\keV} threshold. The majority of this rate results from \Pga-decays in the scintillator. An increase in threshold to \SI{200}{\keV} will remove pulses from low-energy decays such as those from \ICof, \IGdoFt, and \ISmofS and results in a predicted rate of \SI{56.8+-5.5}{\Hz} which is suitably low for use in the LZ OD. It was noted that delayed signals from \Pga events will contribute less to the overall false veto probability of the OD through the use of PSD techniques. Improved cleanliness procedures and more aggressive purification of the Gd compound will also help to lower the rate further.


\section{Acknowledgments}
The authors would like to acknowledge contributions to this work from the following groups within the LZ collaboration: Brown University for providing the R11410-20 PMTs, University of California-Davis for providing amplifiers, and University of Rochester for providing the SkuTek DDC-10 digitizer. We also wish to acknowledge S.A. Hertel for use of the thoron source and V.A. Kudryavtsev for estimates of the expected $^{7}\textrm{Be}$ concentration.

This work was supported by the U.S. Department of Energy (DOE) under award numbers DE-SC0011702, DE-AC02-05CH11231, and DE-SC0012704.

We acknowledge many types of support provided to us by the South Dakota Science and Technology Authority (SDSTA), which developed the Sanford Underground Research Facility (SURF) with an important philanthropic donation from T. Denny Sanford as well as support from the State of South Dakota. SURF is operated by the SDSTA under contract to the Fermi National Accelerator Laboratory for the DOE, Office of Science.

\newpage

\section*{References}
\bibliographystyle{elsarticle-num}
\bibliography{screener,LZ_screener}

\end{document}

%% file: LZPaperSupport/packages.tex
\usepackage{siunitx} 
\usepackage{./LZPaperSupport/hepparticles} 
\usepackage[italic]{heppennames}

\usepackage{mathptmx}  

\usepackage[dvipsnames]{xcolor}
\usepackage{colortbl}

\usepackage{lineno,hyperref}